\documentclass[12pt]{amsart}

\usepackage{epsfig}

\begin{document}

\title[]{\bf Conformal Scalar Propagation inside the
Schwarzschild Black Hole}

\author[]{George Tsoupros \\
       {\em Private address - Beijing}\\
       {\em People's republic of China}\\
}
\thanks{present e-mail address: landaughost@hotmail.com}

\begin{abstract}

The analytic expression obtained in the preceding project for the
massless conformal scalar propagator in the Hartle-Hawking vacuum
state for small values of the Schwarzschild radial coordinate above
$ r = 2M$ is analytically extended into the interior of the
Schwarzschild black hole. The result of the analytical extension
coincides with the exact propagator for a small range of values of
the Schwarzschild radial coordinate below $ r = 2M$ and is an
analytic expression which manifestly features its dependence on the
background space-time geometry. This feature as well as the absence
of any assumptions and prerequisites in the derivation render this
Hartle-Hawking scalar propagator in the interior of the
Schwarzschild black-hole geometry distinct from previous results.
The two propagators obtained in the interior and in the exterior
region of the Schwarzschild black hole are matched across the event
horizon. The result of that match is a massless conformal scalar
propagator in the Hartle-Hawking vacuum state which is shown to
describe particle production by the Schwarzschild black hole.

\end{abstract}

\maketitle

\hspace{3in} \emph{``The future is not what it used to be!"}

\hspace{3in} \emph{From Alan Parker's film ``Angel Heart"}

{\bf I. Introduction}\\

The intimate relation between the Feynman propagator and the
expectation value $ <T_{\mu\nu}>$ of the stress tensor in any
specific vacuum state is the primary reason that the former is an
essential mathematical construct for the study of vacuum activity on
curved manifolds. In a black-hole space-time no less important a
reason is the relation between higher-order radiative effects and
the radiation emitted by black holes in certain vacuum states.
Following the construction of a general expression for the Feynman
Green function in the exterior region of the Schwarzschild black
hole \cite{Candelas} Candelas and Jensen extended that result with
Hartle-Hawking boundary conditions in the interior region of that
space-time \cite{CanJen}. As is the case in \cite{Candelas} the
exact dependence of the Green function on the Schwarzschild radial
variable in \cite{CanJen} remains unknown. Yet another approach to
the Feynman Green function in the interior of the Schwarzschild
black hole has been proposed in \cite{YajNar}. That Green function
is essentially coordinate dependent. Its construction exploits CPT
invariance in order to speculate on a possible vacuum state in the
interior which conforms with the Boulware vacuum state.

In \cite{George} an analytic expression for the conformal scalar
propagator in the Hartle-Hawking vacuum state was provided. That
expression features an explicit dependence on the Schwarzschild
black-hole space-time and is a valid approximation to the unknown
exact propagator in the exterior region of that space-time for small
values of the radial coordinate above $ r = 2M$. This project has
two objectives. The first is to extend that analytic expression for
the Feynman Green function in the interior of the Schwarzschild
black hole. Following that extension the second objective is to
match the scalar propagators in the interior region and in the
exterior region across the event horizon of the Schwarzschild black
hole and to establish that the scalar propagator which emerges as a
result of that match necessarily describes particle production by
the Schwarzschild black hole, in conformity with Hawking radiation.
The analytic expression in which the ensuing analysis will eventuate
will, likewise, be shown to be a valid approximation to the exact
propagator in the interior of the Schwarzschild black hole for a
certain range of radii $ r < 2M$. As was also the case in
\cite{George} an essential advantage of the propagator in which the
analysis herein will eventuate is that it renders the short-distance
behaviour and the singularity structure of the propagator manifest.
In effect, that Green function is especially suited for an analytic
evaluation of $ <T_{\mu\nu}>$ and of $ <\phi^2(x)>$  in the interior
region of the Schwarzschild black hole as well as for a higher-order
perturbative evaluation of the dynamical behaviour of the conformal
scalar field in the presence of interactions in the Schwarzschild
black-hole space-time.

{\bf II. The Hartle-Hawking Green function}\\

The Schwarzschild metric is

\begin{equation}
ds^2 = -(1 - \frac{2M}{r})dt^2 + (1 - \frac{2M}{r})^{-1}dr^2 +
r^2(d\theta^2 + sin^2\theta d\phi^2)
\end{equation}
The analytical extension $ \tau = + it$ of the real-time coordinate
$ t$ in imaginary values results in a Euclidean, positive definite
metric for $ r > 2M$. The apparent singularity which persists at $ r
= 2M$ can be removed by introducing the new radial coordinate
\cite{SHawking}

\begin{equation}
\rho = 4M(1 - \frac{2M}{r})^{\frac{1}{2}}
\end{equation}

Upon replacing

\begin{equation}
\beta = 4M
\end{equation}
the metric in the new coordinates is

\begin{equation}
ds^2 = \rho^2(\frac{1}{\beta^2})d\tau^2 +
(\frac{4r^2}{\beta^2})^2d\rho^2 + r^2(d\theta^2 + sin^2\theta
d\phi^2)
\end{equation}

The coordinate singularity at $ r=2M$ corresponds to the origin $
\rho = 0$ of polar coordinates and is removed by identifying $
\frac{\tau}{4M}$ with an angular coordinate of period $ 2\pi$. In
addition, although the curvature singularity at $ r=0$ can not be
removed by coordinate transformations this procedure can be seen to
avoid it altogether along with the entire $ r < 2M$ region of the
Schwarzschild geometry in real time. This procedure results,
therefore, in a complete singularity-free positive definite,
Euclidean metric which is periodic in the imaginary-time coordinate
$ \tau$. The period $ 8\pi M$ of the imaginary-time coordinate $
\tau$ is the underlying cause of the thermal radiation at
temperature $ T = (8\pi M)^{-1}$ emitted by the Schwarzschild black
hole.

In resulting in a singularity-free positive definite, Euclidean
metric for $ r \geq 2M$ this procedure also singles out in that
region the Hartle-Hawking Euclidean Green function for the massless
scalar field as that solution to \footnote{where for the conformally
invariant theory it is $ \xi = \frac{1}{6}$}

\begin{equation}
[- \square_{x_2} + \xi R(x_2)]D(x_1, x_2) = \delta(x_1, x_2)
\end{equation}
which is uniquely specified by the requirement that (i) $ D(x_2 -
x_1)$ be regular on the Euclidean section of (1) - and, thereby,
periodic in the imaginary-time coordinate $ \tau$ - and (ii) vanish
as $ r \rightarrow \infty$ - and, consequently, as $ \rho
\rightarrow \beta$. By imposing the requirement $ \rho^2 << \beta^2$
the solution to (5) was, in \cite{George}, determined to be

$$
D(x_2 - x_1) =
$$

$$
\frac{2}{\beta}\frac{1}{\sqrt{\rho_1\rho_2}}\sum_{l=0}^{\infty}\sum_{m=-l}^{l}Y_{lm}(\theta_2,
\phi_2)Y_{lm}^*(\theta_1, \phi_1)\sum_{p =
0}^{\infty}e^{i\frac{p}{\beta}(\tau_2 -
\tau_1)}\int_{u_0[p]}^{\infty}du\frac{cos[\frac{\pi}{4\beta}(4u + 2p
+ 3)(\rho_2 - \rho_1)]}{\pi^2u^2 + 4(l^2 + l + 1)}
$$

$$
- \frac{2}{\beta^{\frac{3}{2}}}\frac{1}{\sqrt{\rho_1}}\times
$$

$$
\sum_{l = 0} ^{\infty}\sum_{m = -l}^{l}\sum_{p =
0}^{\infty}\int_{u_0'[p]}^{\infty}du\frac{cos[\frac{\pi}{4\beta}(4u
+ 2p + 3)(\beta - \rho_1)]}{\pi^2u^2 + 4(l^2 + l +
1)}\frac{J_p(\frac{2i}{\beta}\sqrt{l^2 + l +
1}\rho_2)}{J_p(2i\sqrt{l^2 + l + 1})}e^{i\frac{p}{\beta}(\tau_2 -
\tau_1)}Y_{lm}(\theta_2, \phi_2)Y_{lm}^*(\theta_1, \phi_1) ~~~;~~
$$

\begin{equation}
u_0 >> p ~~~~;~~~ u_0' >> p ~~~;~~~ \frac{\pi u}{\beta}\rho_{2,1}
>> p
\end{equation}
with the range of validity for the Green function in (6) specified
by

\begin{equation}
0 \leq \rho_i^2 \leq \frac{\beta^2}{100} ~~~~ ; ~~~~ i = 1, 2
\end{equation}
or, equivalently, by \footnote{in \cite{George} the upper bound was
inadvertently stated to be 2.0050M instead of 2.050M. The correct
upper bound 2.050M increases the already large range of validity in
(6) by yet one more order of magnitude.}

\begin{equation}
2M \leq r \leq 2.050M
\end{equation}

It is now desirable to, analytically, extend the Euclidean Green
function in (6) to the interior region $ r < 2M$ of (1). In this
region it is not sufficient to pass to imaginary time as that would
entail a metric with signature (-, -, +, +) \cite{CanJen}. It is
therefore necessary to implement the additional change $ \theta
\rightarrow i\tilde{\theta}$ as a result of which (1) becomes

\begin{equation}
ds^2 = -\big{[}[\frac{2M}{r} -  1]d\tau^2 + \frac{1}{\frac{2M}{r} -
1}dr^2 + r^2(d\tilde{\theta}^2 + sinh^2\tilde{\theta}d\phi^2)\big{]}
\end{equation}
This metric can be seen to have a negative-definite signature for $
r < 2M$. The reasons that the Euclidean Green function subjected to
the Hartle-Hawking requirement of periodicity in imaginary time can,
in the interior region, only be obtained through an analytical
extension of the corresponding Green function in the exterior region
of the black-hole geometry - as opposed to, through solving (5) for
$ 0 < r \leq 2M$ - have been sufficiently analyzed in \cite{CanJen}.
As stated, the exact dependence which the Green function in
\cite{CanJen} has on the radial variable is unknown. By contrast,
the dependence of the Green function in (6) on space-time is
explicit. Therefore an appropriate analytical extension of (6) to
values of $ D(x_2 - x_1)$ corresponding to $ 0 < r \leq 2M$ is
expected to yield a well-defined and unique propagator in the
interior region of the Schwarzschild black hole. Such a propagator
will identically satisfy two central demands. These are, the demand
for the correct divergence structure at the coincidence space-time
limit $ x_2 \rightarrow x_1$ and the demand for regularity away from
that limit. The result of the analytical extension of (6) must, in
the interior of the Schwarzschild black hole, define a Green
function whose singularity structure will correspond to a quadratic
divergence and which, away from the coincidence space-time limit,
will be a regular (that is, single valued and analytic) function of
space-time for all values of the space-time coordinates in its
domain. In what follows the appropriate analytical extension of (6)
will, indeed, yield such a Green function.

{\bf III. The Analytical Extension - Singular Part}\\

The singular part $ D_{as}(x_2 - x_1)$ of the Green function in (6)
- that is, the part which is singular at $ x_2 \rightarrow x_1$ and
coincides with the asymptotic expression of that Green function
\cite{George} - is

$$
D_{as}(x_2 - x_1)=
$$

\begin{equation}
\frac{2}{\beta}\frac{1}{\sqrt{\rho_1\rho_2}}\sum_{l=0}^{\infty}\sum_{m=-l}^{l}Y_{lm}(\theta_2,
\phi_2)Y_{lm}^*(\theta_1, \phi_1)\sum_{p =
0}^{\infty}e^{i\frac{p}{\beta}(\tau_2 -
\tau_1)}\int_{u_0[p]}^{\infty}du\frac{cos[\frac{\pi}{4\beta}(4u + 2p
+ 3)(\rho_2 - \rho_1)]}{\pi^2u^2 + 4(l^2 + l + 1)}
\end{equation}
the radial-temporal sector of which is the sum of

\begin{equation}
\frac{e^{i\frac{3\pi}{4\beta}(\rho_2 -
\rho_1)}}{\beta\sqrt{\rho_1\rho_2}}\sum_{p =
0}^{\infty}e^{i\frac{p}{\beta}(\tau_2 -
\tau_1)}e^{i\frac{\pi}{2\beta}p(\rho_2 -
\rho_1)}\int_{u_0[p]}^{\infty}du\frac{e^{i\frac{\pi}{\beta}u(\rho_2
- \rho_1)}}{\pi^2u^2 + 4(l^2 + l + 1)}
\end{equation}

and

\begin{equation}
\frac{e^{-i\frac{3\pi}{4\beta}(\rho_2 -
\rho_1)}}{\beta\sqrt{\rho_1\rho_2}}\sum_{p =
0}^{\infty}e^{i\frac{p}{\beta}(\tau_2 -
\tau_1)}e^{-i\frac{\pi}{2\beta}p(\rho_2 -
\rho_1)}\int_{u_0[p]}^{\infty}du\frac{e^{-i\frac{\pi}{\beta}u(\rho_2
- \rho_1)}}{\pi^2u^2 + 4(l^2 + l + 1)}
\end{equation}

These two expressions are, of course, valid in the exterior region
of the Schwarzschild black-hole space-time - that is, in the region
defined by $ r \geq 2M$ and, for that matter, by $ \rho \geq 0$. The
naive replacement of that range in these two expressions by $ 0 \leq
r < 2M$ with its concomitant imaginary values of $ \rho$ causes
either one of them to be sharply divergent for arbitrary space-time
separations in the interior region. Such a replacement is,
consequently, unacceptable. Clearly, what is required in the
interior is a mathematical expression for the propagator the radial
sector of which will exclusively feature purely imaginary values of
$ \rho$ while remaining convergent away from the coincidence
space-time limit $ x_2 \rightarrow x_1$. In what follows, both (11)
and (12) will be respectively subjected to a specific analytical
extension which will be shown to correspond identically to the
analytical extension of (11) and (12) in the interior region of the
black hole. In both (11) and (12) the assumption is that $ \rho_2 -
\rho_1
> 0$ and $ \tau_2 - \tau_1 > 0$. The analytical extension will first
be applied to (11).

Upon extending $ u \in R$ to $ \tilde{u} \in C$ the function which
is integrated over $ u$ develops poles at $ \pm i\frac{ \sqrt{4(l^2
+ l + 1)}}{\pi}$. Hence, over the closed contour $ c_{\tilde{u}}$ of
Fig. 1 - specified by the segments (i) $ u - u_0 > 0$ (ii) $
|\tilde{u} - u_0|e^{i\theta_{\tilde{u}}'}$, $ 0 \leq
\theta_{\tilde{u}}' \leq \frac{\pi}{2}$, $ |\tilde{u}| = u$, and
(iii) $ |\tilde{u} - u_0|e^{i\frac{\pi}{2}}$ - by Cauchy's theorem
it is

\begin{equation}
\oint_{c_{\tilde{u}}}d\tilde{u}\frac{e^{i\frac{\pi}{\beta}\tilde{u}(\rho_2
- \rho_1)}}{\pi^2\tilde{u}^2 + 4(l^2 + l + 1)} = 0
\end{equation}

\begin{figure}[h]
\centering\epsfig{figure=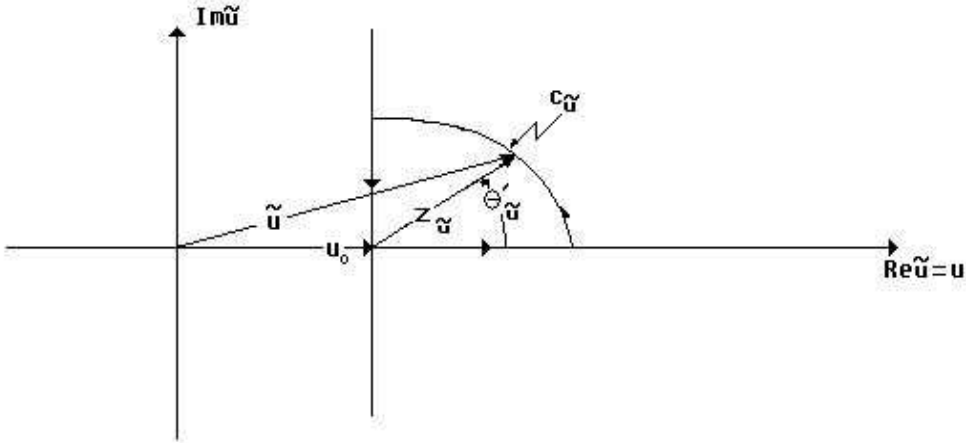, height =85mm,width=141mm}
\caption{Integration over the infinite contour $ c_{\tilde{u}}$ in
the $ \tilde{u}$-complex plane. The choice of $ c_{\tilde{u}}$ is
dictated by the demand for vanishing contribution along the infinite
quadrant.}
\end{figure}

Setting $ \tilde{u} - u_0 = z_{\tilde{u}}$ it is on the large
quadrant $ c_{\tilde{u}}^{\theta'}$ of $ c_{\tilde{u}}$

$$
I_{c_{\tilde{u}}^{\theta'}} =
\int_{c_{\tilde{u}}^{\theta'}}d\tilde{u}\frac{e^{i\frac{\pi}{\beta}\tilde{u}(\rho_2
- \rho_1)}}{\pi^2\tilde{u}^2 + 4(l^2 + l + 1)} =
e^{i\frac{\pi}{\beta}u_0[p](\rho_2 - \rho_1)}
\int_{c_{\tilde{u}}^{\theta'}}dz_{\tilde{u}}\frac{e^{i\frac{\pi}{\beta}z_{\tilde{u}}(\rho_2
- \rho_1)}}{\pi^2(z_{\tilde{u}} + u_0)^2 + 4(l^2 + l + 1)} =
$$

$$
ie^{i\frac{\pi}{\beta}u_0[p](\rho_2 -
\rho_1)}\int_0^{\frac{\pi}{2}}d\theta'_{\tilde{u}}|z_{\tilde{u}}|
e^{i\theta_{\tilde{u}}'}\frac{e^{i\frac{\pi}{\beta}|z_{\tilde{u}}|
cos\theta_{\tilde{u}}'(\rho_2 -
\rho_1)}e^{-\frac{\pi}{\beta}|z_{\tilde{u}}|sin\theta_{\tilde{u}}'(\rho_2
- \rho_1)}}{\pi^2(z_{\tilde{u}} + u_0)^2 + 4(l^2 + l + 1)}
$$
so that letting $ |\tilde{u}| \rightarrow \infty$ while keeping $
u_0$ at a fixed value it is

\begin{equation}
lim_{|z_{\tilde{u}}| \rightarrow \infty}I_{c_{\tilde{u}}^{\theta'}}
= 0
\end{equation}

The contribution along the stated third segment of the infinite
countour $ c_{\tilde{u}}$ is

$$
I_{c_{\tilde{u}}^{\pi/2}} = \int_{u_0 +
i\infty}^{u_0[p]}d\tilde{u}\frac{e^{i\frac{\pi}{\beta}\tilde{u}(\rho_2
- \rho_1)}}{\pi^2\tilde{u}^2 + 4(l^2 + l + 1)} =
e^{i\frac{\pi}{\beta}u_0[p](\rho_2 - \rho_1)}
\int_{i\infty}^0dz_{\tilde{u}}\frac{e^{i\frac{\pi}{\beta}z_{\tilde{u}}(\rho_2
- \rho_1)}}{\pi^2(z_{\tilde{u}} + u_0)^2 + 4(l^2 + l + 1)}
$$
with $ |z_{\tilde{u}}| = w$ it is

\begin{equation}
I_{c_{\tilde{u}}^{\pi/2}} = e^{i\frac{\pi}{\beta}u_0[p](\rho_2 -
\rho_1)}
\int_{\infty}^0dwe^{i\frac{\pi}{2}}\frac{e^{-\frac{\pi}{\beta}w(\rho_2
- \rho_1)}}{\pi^2(u_0^2 - w^2) + 4(l^2 + l + 1) + 2i\pi^2u_0w}
\end{equation}

As a consequence of (14) and (15) the statement in (13) implies

\begin{equation}
\int_{u_0[p]}^{\infty}du\frac{e^{i\frac{\pi}{\beta}u(\rho_2 -
\rho_1)}}{\pi^2u^2 + 4(l^2 + l + 1)} =
ie^{i\frac{\pi}{\beta}u_0[p](\rho_2 - \rho_1)}
\int_{0}^{\infty}dw\frac{e^{-\frac{\pi}{\beta}w(\rho_2 -
\rho_1)}}{\pi^2(u_0^2 - w^2) + 4(l^2 + l + 1) + 2i\pi^2u_0w}
\end{equation}

In view of this result (11) becomes

$$
i\frac{e^{i\frac{3\pi}{4\beta}(\rho_2 -
\rho_1)}}{\beta\sqrt{\rho_1\rho_2}}\sum_{p =
0}^{\infty}e^{i\frac{p}{\beta}(\tau_2 -
\tau_1)}e^{i\frac{\pi}{2\beta}p(\rho_2 -
\rho_1)}e^{i\frac{\pi}{\beta}u_0[p](\rho_2 - \rho_1)}\times
$$

\begin{equation}
\int_{0}^{\infty}dw\frac{e^{-\frac{\pi}{\beta}w(\rho_2 -
\rho_1)}}{\pi^2(u_0^2[p] - w^2) + 4(l^2 + l + 1) + 2i\pi^2u_0[p]w}
\end{equation}
In order to advance the analytical extension of (11) it is necessary
to convert the infinite series over $ p$ in (17) to a contour
integral. Since, upon extending $ p \in R$ to $ \tilde{p} \in C$,
the function of $ \tilde{p}$ which emerges from (17) tends to zero
at $ |\tilde{p}| \rightarrow \infty$ faster than $ |\tilde{p}|^{-1}$
the residue theorem implies that (17) can be expressed as

$$
\frac{i}{\beta}\frac{1}{2i}\oint_{c_{\tilde{p}}}d\tilde{p}\frac{(-1)^{\tilde{p}}}{sin{(\pi\tilde{p})}}
e^{i\frac{\tilde{p}}{\beta}(\tau_2 -
\tau_1)}e^{i\frac{\pi}{2\beta}\tilde{p}(\rho_2 -
\rho_1)}e^{i\frac{\pi}{\beta}u_0[\tilde{p}](\rho_2 -
\rho_1)}\frac{e^{i\frac{3\pi}{4\beta}(\rho_2 -
\rho_1)}}{\sqrt{\rho_1\rho_2}}\times
$$

\begin{equation}
\int_{0}^{\infty}dw\frac{e^{-\frac{\pi}{\beta}w(\rho_2 -
\rho_1)}}{\pi^2(u_0^2[\tilde{p}] - w^2) + 4(l^2 + l + 1) +
2i\pi^2u_0[\tilde{p}]w}
\end{equation}
on condition that the infinite contour $ c_{\tilde{p}}$ in the $
\tilde{p}$-complex plane encompasses all non-negative integers, as
in Fig.2. The reason the factor $
\frac{e^{i\frac{3\pi}{4\beta}(\rho_2 -
\rho_1)}}{\sqrt{\rho_1\rho_2}}$ appears in the integrand will be
elucidated in what follows.

\begin{figure}[h]
\centering\epsfig{figure=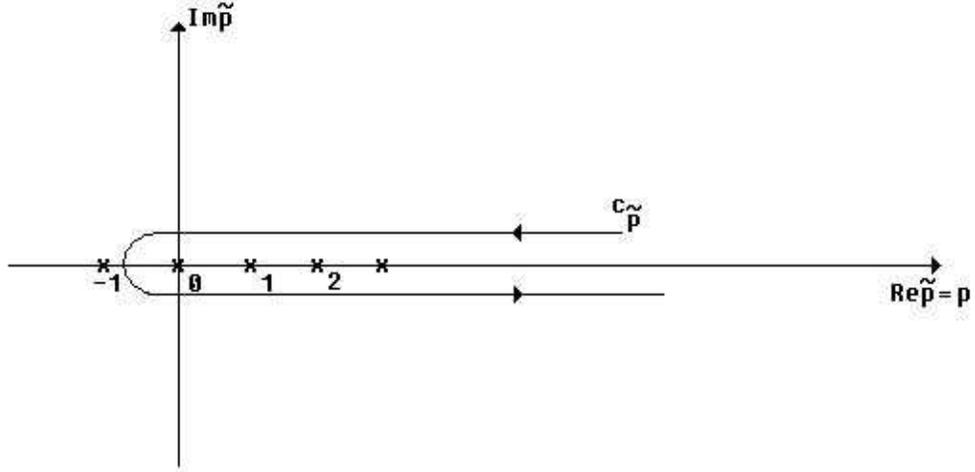, height =85mm,width=141mm}
\caption{An infinite contour for the integral in (18)}
\end{figure}

Attention is invited to the fact that since the expression $
\frac{\tau_2 - \tau_1}{\beta}$ is an angular coordinate the presence
of $ \tilde{p} \in C$ in the corresponding exponential in (18)
imposes the physical requirement that the evaluation of the contour
integral in the latter yield a single-valued function of the
temporal difference $ \tau_2 - \tau_1$. That this is, indeed, the
case follows from the character of (18) as an integral
representation of (11).

The integral in (18) taken along the contour of Fig.2 can not, in
any physically meaningful way, be associated with the imaginary
values of $ \rho$ stemming from (2) for $ r < 2M$. The primary
reason is that, on general theoretical grounds, any analytical
extension in transfer space necessarily induces a corresponding
extension in the coordinate variables which precludes the arbitrary
replacement of $ \rho$ by $ \pm i|\rho|$. In addition, the arbitrary
replacement of $ \rho$ by $ -i|\rho|$ in (18) would generate an
oscillatory integrand of a sharply divergent amplitude at $
|\tilde{p}| \rightarrow \infty$ \footnote{The arbitrary replacement
$ \rho \rightarrow +i|\rho|$ would, likewise, generate the same
effect to the corresponding analytical extension of (12).}. However,
the merit of (18) lies in the fact that the contour $ c_{\tilde{p}}$
in Fig.2 can be properly deformed so as to yield a convergent
expression in the interior of the black hole. Since integration in
(18) along any contour which includes identically the same infinite
set of poles as that in Fig.2 results invariably in (17) the contour
in Fig.2 can be deformed to one which (i) avoids each pole through
an arbitrarily small semi-circle just below the real $ p$-axis, (ii)
extends the path of integration in the $ \tilde{p}$-complex plane
from the real axis to the imaginary axis through a large segment of
constant radius and of positive ("counterclockwise") orientation and
(iii) closes the path of integration along the imaginary axis from
positive imaginary infinity to the circumference of a circular
segment of vanishing radius about zero. The corresponding contour $
c'_{\tilde{p}}$ is depicted in Fig.3.

\begin{figure}[h]
\centering\epsfig{figure=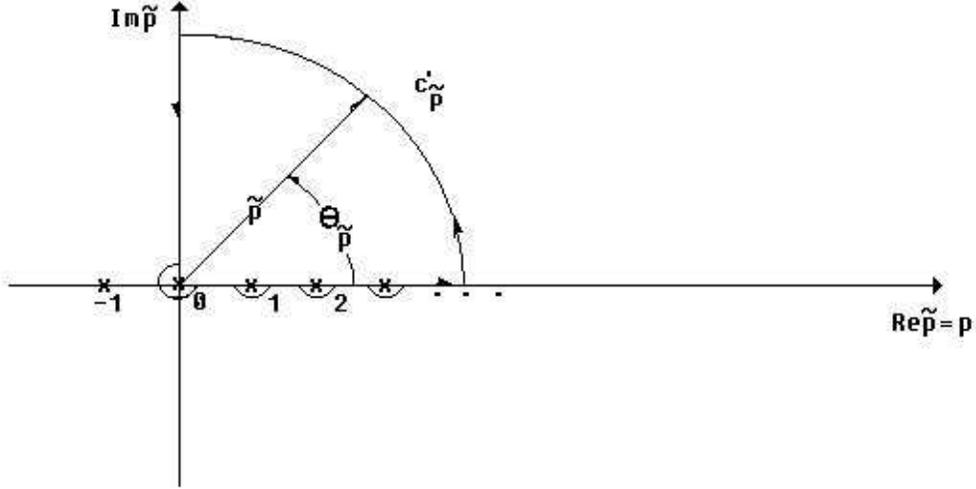, height
=85mm,width=141mm} \caption{An equivalent infinite contour for the
integral in (18)}
\end{figure}

The objective of a convergent expression for the propagator in the
interior region necessitates a careful use of the rotation $ p
\rightarrow |\tilde{p}|e^{i\theta_{\tilde{p}}}$ along the stated
contour in Fig.3. First of all, since in view of (6) it is $ u_0[p]
>> p$ the stated rotation implies that $ u_0[|\tilde{p}|]
>> |\tilde{p}|$. At once, this condition is valid for each $
\theta_{\tilde{p}} \in [0, \frac{\pi}{2}]$ in the $
\tilde{p}$-complex plane since the integration along the contour $
c_{\tilde{u}}$ which eventuated in (16) has been performed in the $
\tilde{u}$-complex plane and is, for that matter, irrelevant to $
\theta_{\tilde{p}}$. In effect, the rotation $ p \rightarrow
|\tilde{p}|e^{i\theta_{\tilde{p}}}$ along the contour $
c'_{\tilde{p}}$ implies that

\begin{equation}
u_0[p] \rightarrow |u_0[\tilde{p}]|e^{i\theta_{\tilde{p}}} ~~~ ; ~~~
|u_0[\tilde{p}]| = u_0[|\tilde{p}|] >> |\tilde{p}|
\end{equation}

In addition, since - as stated - the analytical extension $ p
\rightarrow |\tilde{p}|e^{i\theta_{\tilde{p}}}$ must be offset by a
corresponding extension in the associated coordinate variables it is
necessary that the integration along the contour $ c'_{\tilde{p}}$
in Fig.3 be advanced in the context of the simultaneous
substitutions

\begin{equation}
\tau_{2} = e^{-i\theta_{\tilde{p}}}\xi_{2} ~~~ ; ~~~ \tau_{1} =
e^{-i\theta_{\tilde{p}}}\xi_{1} ~~~ ; ~~~ \xi_{2}, \xi_{1} \in R
\end{equation}
and

\begin{equation}
\rho_{2} = e^{-i\theta_{\tilde{p}}}\zeta_{2} ~~~ ; ~~~ \rho_{1} =
e^{-i\theta_{\tilde{p}}}\zeta_{1} ~~~ ; ~~~ \zeta_{2}, \zeta_{1} \in
R
\end{equation}
The new variables $ \xi_{i}$ and $ \zeta_{i} ~~~ ; ~~~ i = 1, 2$
being the corresponding magnitudes of $ \tau_{i}$ and $ \rho_{i}$
respectively coincide with the latter in the exterior region of the
black-hole geometry. The physical significance of this fact will be
discussed in the final result. It is stressed, however, that (21)
reflects a general property of the radial variable $ \rho$ in the
event that $ p \rightarrow |\tilde{p}|e^{i\theta_{\tilde{p}}}$.
Consequently, the factor $ \frac{e^{i\frac{3\pi}{4\beta}(\rho_2 -
\rho_1)}}{\sqrt{\rho_1\rho_2}}$ - which in (11) is $p$-independent -
develops, as a result of (21), an implicit dependence on $
\tilde{p}$ and must be included in the contour integral. This is
explicitly the case in (18).

Deforming the contour $ c_{\tilde{p}}$ to $ c'_{\tilde{p}}$ and
replacing (19), (20) and (21) in the integral over $ \tilde{p}$
featured in (18), now taken over the large quadrant of $
c'_{\tilde{p}}$, results in

$$
I_{c'_{\tilde{p}}}^{\theta_{\tilde{p}}} =
i|\tilde{p}|\int_{0}^{\frac{\pi}{2}}d\theta_{\tilde{p}}e^{i\theta_{\tilde{p}}}\frac{(-1)^{\tilde{p}}}{sin{(\pi\tilde{p})}}
e^{i\frac{|\tilde{p}|}{\beta}(\xi_2 -
\xi_1)}e^{i\frac{\pi}{2\beta}|\tilde{p}|(\zeta_2 -
\zeta_1)}e^{i\frac{\pi}{\beta}u_0[|\tilde{p}|](\zeta_2 -
\zeta_1)}\frac{e^{i\frac{3\pi}{4\beta}e^{-i\theta_{\tilde{p}}}(\zeta_2
- \zeta_1)}}{e^{-i\theta_{\tilde{p}}}\sqrt{\zeta_1\zeta_2}}\times
$$

\begin{equation}
\int_{0}^{\infty}dw\frac{e^{-\frac{\pi}{\beta}we^{-i\theta_{\tilde{p}}}(\zeta_2
- \zeta_1)}}{\pi^2(u_0^2[|\tilde{p}|]e^{i2\theta_{\tilde{p}}} - w^2)
+ 4(l^2 + l + 1) + 2i\pi^2u_0[|\tilde{p}|]e^{i\theta_{\tilde{p}}}w}
\end{equation}

It can be seen at this stage that (11) admits, in principle, several
extensions depending on the manner in which $ (-1)^{\tilde{p}}$ is
expressed. Apparently, certain conditions are necessary in order to
select a unique, physically valid analytical extension. To that
effect it must be observed that the Green function in (6) features,
in its singular part at the coincidence space-time limit $ x_2
\rightarrow x_1$, a logarithmic divergence in $ 1 + 1$-dimensions
and a quadratic divergence in four dimensions \cite{George}. Such a
divergence structure identically characterises also (11) and (12).
It is imperative, for that matter, that the same divergence
structure characterise the analytical extension of (11) and (12)
independently since that extension is, essentially, an integral
representation of (11) and (12) respectively. In addition, the
stated divergence structure is, on general theoretical grounds
\cite{George}, also expected in the interior of the black hole - on
condition that the Schwarzschild radial coordinates associated with
the Feynman propagator receive values well above the $ r = 0$,
singularity-related value. As a consequence, any analytical
extension of these two expressions which does not preserve their
stated divergence structure must be ruled out. At once, consistency
with this requirement imposes the additional demand that the stated
divergence structure be preserved at each stage of the analytical
extension of (11) and, independently, of (12).

As a consequence of this demand it is, in (22), $ (-1)^{\tilde{p}} =
(e^{-i\pi})^{\tilde{p}}$ so that

$$
\frac{(-1)^{\tilde{p}}}{sin(\pi\tilde{p})} =
$$

$$
e^{-i\pi|\tilde{p}|cos\theta_{\tilde{p}}}\frac{2e^{\pi|\tilde{p}|sin\theta_{\tilde{p}}}}{sin(\pi|\tilde{p}|cos\theta_{\tilde{p}})
(e^{\pi|\tilde{p}|sin\theta_{\tilde{p}}} +
e^{-\pi|\tilde{p}|sin\theta_{\tilde{p}}}) +
icos(\pi|\tilde{p}|cos\theta_{\tilde{p}})(e^{\pi|\tilde{p}|sin\theta_{\tilde{p}}}
- e^{-\pi|\tilde{p}|sin\theta_{\tilde{p}}})}
$$

The right side of this relation can, for large values of $
|\tilde{p}|$, be approximated by

$$
e^{-i\pi|\tilde{p}|cos\theta_{\tilde{p}}}\frac{2e^{\pi|\tilde{p}|sin\theta_{\tilde{p}}}}{[sin(\pi|\tilde{p}|cos\theta_{\tilde{p}})
+
icos(\pi|\tilde{p}|cos\theta_{\tilde{p}})]e^{\pi|\tilde{p}|sin\theta_{\tilde{p}}}}
= -2i
$$
in view of the fact that along the quadrant of $ c'_{\tilde{p}}$ it
is $ 0 \leq sin\theta_{\tilde{p}} \leq 1$.

Replacing this result in (22) it is seen that as $ |\tilde{p}|
\rightarrow \infty$ the expression $
I_{c'_{\tilde{p}}}^{\theta_{\tilde{p}}}$ is purely oscillatory in
the ultra-violet domain of the integral over $ w$, if $ x_2 \neq
x_1$. In order to ensure a vanishing result at $ |\tilde{p}|
\rightarrow \infty$, for that matter, it is also necessary to
implement the substitution

$$
\xi_2 - \xi_1 \rightarrow (\xi_2 - \xi_1) + iv ~~~~ ; ~~~~ 0 < v <<
1 ~~~~ ; ~~~~ x_2 \neq x_1 ~~~~ ; ~~~~ \theta_{\tilde{p}} \in (0,
\frac{\pi}{2})
$$
on the understanding that the limit $ v \rightarrow 0$ is implicit
at the end of the calculation in this range of $ \theta_{\tilde{p}}$
and, independently, at $ x_2 \rightarrow x_1$ prior to taking $
|\tilde{p}| \rightarrow \infty$.

In effect, it is

\begin{equation}
lim_{|\tilde{p}| \rightarrow \infty}
I_{c'_{\tilde{p}}}^{\theta_{\tilde{p}}} = 0 ~~~~;~~~~ x_2 \neq x_1
\end{equation}
At once, it can be seen that taking the coincidence space-time limit
$ x_2 \rightarrow x_1$ prior to taking $ |\tilde{p}| \rightarrow
\infty$ results in the desired logarithmic divergence, stated above.
In this respect it can readily be confirmed that the replacement $
(-1)^{\tilde{p}} = (e^{i\pi})^{\tilde{p}}$ is inadmissible in (22)
as, at $ |\tilde{p}| \rightarrow \infty$, it would yield a vanishing
result also for $ x_2 \rightarrow x_1$.

The integral over $ \tilde{p}$ in (18), now taken along that segment
of $ c'_{\tilde{p}}$ which lies on the imaginary axis, corresponds
to $ \theta_{\tilde{p}} = \frac{\pi}{2}$. Again, in view of (19),
(20) and (21) that integral is

$$
I_{c'_{\tilde{p}}}^{\frac{\pi}{2}} = \int_{\infty}^{\epsilon
\rightarrow
0}d|\tilde{p}|i\frac{(-1)^{i|\tilde{p}|}}{isinh{(\pi|\tilde{p}|)}}
e^{i\frac{|\tilde{p}|}{\beta}(\xi_2 -
\xi_1)}e^{i\frac{\pi}{2\beta}|\tilde{p}|(\zeta_2 -
\zeta_1)}e^{i\frac{\pi}{\beta}u_0[|\tilde{p}|](\zeta_2 -
\zeta_1)}\frac{e^{i\frac{3\pi}{4\beta}(-i)(\zeta_2 -
\zeta_1)}}{(-i)\sqrt{\zeta_1\zeta_2}}\times
$$

\begin{equation}
\int_{0}^{\infty}dw\frac{e^{-\frac{\pi}{\beta}w(-i)(\zeta_2 -
\zeta_1)}}{-\pi^2(u_0^2[|\tilde{p}|] + w^2) + 4(l^2 + l + 1) -
2\pi^2u_0[|\tilde{p}|]w}
\end{equation}
where - as is evident in Fig.3 - $ \epsilon << 1$ corresponds to the
radius of the circular segment centred at zero.

Replacing

$$
\frac{(-1)^{i|\tilde{p}|}}{sinh{(\pi|\tilde{p}|)}} =
2\frac{e^{\pi|\tilde{p}|}}{e^{\pi|\tilde{p}|} - e^{-\pi|\tilde{p}|}}
= \frac{2}{1 - e^{-2\pi|\tilde{p}|}}
$$
in (24) yields

$$
I_{c'_{\tilde{p}}}^{\frac{\pi}{2}} = 2i\int_{\infty}^{\epsilon
\rightarrow 0}d|\tilde{p}|\frac{1}{1 - e^{-2\pi|\tilde{p}|}}
e^{i\frac{|\tilde{p}|}{\beta}(\xi_2 -
\xi_1)}e^{i\frac{\pi}{2\beta}|\tilde{p}|(\zeta_2 -
\zeta_1)}e^{i\frac{\pi}{\beta}u_0[|\tilde{p}|](\zeta_2 -
\zeta_1)}\frac{e^{\frac{3\pi}{4\beta}(\zeta_2 -
\zeta_1)}}{\sqrt{\zeta_1\zeta_2}}\times
$$

\begin{equation}
\int_{0}^{\infty}dw\frac{e^{i\frac{\pi}{\beta}w(\zeta_2 -
\zeta_1)}}{-\pi^2(u_0^2[|\tilde{p}|] + w^2) + 4(l^2 + l + 1) -
2\pi^2u_0[|\tilde{p}|]w}
\end{equation}
Again, at $ x_2 \rightarrow x_1$ this expression manifestly features
the desired logarithmic divergence. If, instead, the analytical
extension were based on $ (-1)^{\tilde{p}} = (e^{i\pi})^{\tilde{p}}$
then (25) would vanish at $ x_2 \rightarrow x_1$.

In what follows the segment of the contour $ c'_{\tilde{p}}$ in
Fig.3 which extends along $ Re\tilde{p}$ obviating each pole other
than $ \tilde{p} = 0$ through an arbitrarily small semicircle and
the pole at $ \tilde{p} = 0$ through an arbitrarily small circular
quadrant will be denoted by $ c_{\tilde{p}}^{+}$. For notational
convenience the corresponding Cauchy principal value, that is the
integral over $ \tilde{p}$ in (18) evaluated along $
c_{\tilde{p}}^{+}$, will be denoted by $ I_{c'_{\tilde{p}}}^{+}$. As
a result of (23) it follows that the integral over $ \tilde{p}$ in
(18) taken along the infinite contour $ c'_{\tilde{p}}$ is equal to
$ I_{c'_{\tilde{p}}}^{+} + I_{c'_{\tilde{p}}}^{\frac{\pi}{2}} +
I_{c'_{\tilde{p}}}^{sc}$, where $ I_{c'_{\tilde{p}}}^{sc}$ is the
contribution along the small semi-circle centred at zero and
extending from $ \theta_{\tilde{p}} = \frac{\pi}{2}$ to $
\theta_{\tilde{p}} = \frac{3\pi}{2}$ in Fig.3. In order to evaluate
$ I_{c'_{\tilde{p}}}^{+}$ the integral over $ \tilde{p}$ in (18)
will now be evaluated along the closed contour $
\tilde{c}_{\tilde{p}}$ which (i) extends along $ c_{\tilde{p}}^{+}$,
(ii) extends the path of integration in the $ \tilde{p}$-complex
plane from the real axis to the imaginary axis through a large
segment of constant radius which is, this time, of negative
("clockwise") orientation below the real axis and (iii) closes the
path of integration along the imaginary axis from negative imaginary
infinity to the circumference of the circular quadrant of vanishing
radius about zero. The contour $ \tilde{c}_{\tilde{p}}$ is depicted
in Fig.4.

\begin{figure}[h]
\centering\epsfig{figure=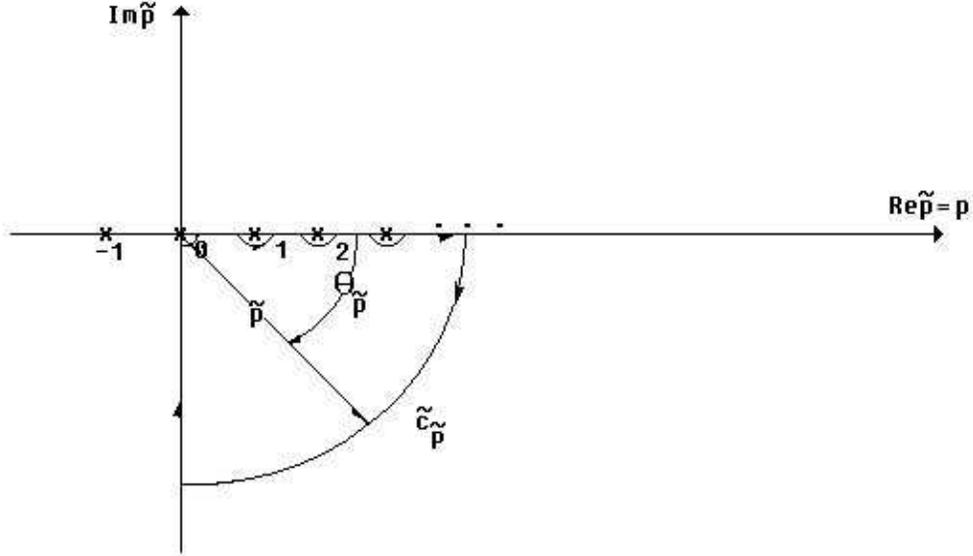, height
=85mm,width=141mm} \caption{The infinite contour $
\tilde{c}_{\tilde{p}}$ in the $ \tilde{p}$-complex plane. The pole
at $ \tilde{p} = 0$ is obviated "counterclockwise" through a
quadrant of vanishing radius. The horizontal segment of $
\tilde{c}_{\tilde{p}}$, including that quadrant, is $
c_{\tilde{p}}^{+}$.}
\end{figure}

Replacing again (19), (20) and (21) in the integral over $
\tilde{p}$ featured in (18), now taken over the large quadrant of $
\tilde{c}_{\tilde{p}}$, results in

$$
I_{\tilde{c}_{\tilde{p}}}^{\theta_{\tilde{p}}} =
i|p|\int_{0}^{-\frac{\pi}{2}}d\theta_{\tilde{p}}e^{i\theta_{\tilde{p}}}\frac{(-1)^{\tilde{p}}}{sin{(\pi\tilde{p})}}
e^{i\frac{|\tilde{p}|}{\beta}(\xi_2 -
\xi_1)}e^{i\frac{\pi}{2\beta}|\tilde{p}|(\zeta_2 -
\zeta_1)}e^{i\frac{\pi}{\beta}u_0[|\tilde{p}|](\zeta_2 -
\zeta_1)}\frac{e^{i\frac{3\pi}{4\beta}e^{-i\theta_{\tilde{p}}}(\zeta_2
- \zeta_1)}}{e^{-i\theta_{\tilde{p}}}\sqrt{\zeta_1\zeta_2}}\times
$$

\begin{equation}
\int_{0}^{\infty}dw\frac{e^{-\frac{\pi}{\beta}we^{-i\theta_{\tilde{p}}}(\zeta_2
- \zeta_1)}}{\pi^2(u_0^2[|\tilde{p}|]e^{i2\theta_{\tilde{p}}} - w^2)
+ 4(l^2 + l + 1) + 2i\pi^2u_0[|\tilde{p}|]e^{i\theta_{\tilde{p}}}w}
\end{equation}
In this context the demand for a logarithmic divergence at each
stage of the analytical extension as $ x_2 \rightarrow x_1$ implies
that $ (-1)^{\tilde{p}} = (e^{i\pi})^{\tilde{p}}$ so that

$$ \frac{(-1)^{\tilde{p}}}{sin(\pi\tilde{p})} = $$

$$
e^{i\pi|\tilde{p}|cos\theta_{\tilde{p}}}\frac{2e^{-\pi|\tilde{p}|sin\theta_{\tilde{p}}}}{sin(\pi|\tilde{p}|cos\theta_{\tilde{p}})
(e^{\pi|\tilde{p}|sin\theta_{\tilde{p}}} +
e^{-\pi|\tilde{p}|sin\theta_{\tilde{p}}}) +
icos(\pi|\tilde{p}|cos\theta_{\tilde{p}})(e^{\pi|\tilde{p}|sin\theta_{\tilde{p}}}
- e^{-\pi|\tilde{p}|sin\theta_{\tilde{p}}})}
$$

The right side of this relation can, for large values of $
|\tilde{p}|$, be approximated by

$$
e^{i\pi|\tilde{p}|cos\theta_{\tilde{p}}}\frac{2e^{-\pi|\tilde{p}|sin\theta_{\tilde{p}}}}{[sin(\pi|\tilde{p}|cos\theta_{\tilde{p}})
-
icos(\pi|\tilde{p}|cos\theta_{\tilde{p}})]e^{-\pi|\tilde{p}|sin\theta_{\tilde{p}}}}
= 2i
$$
in view of the fact that along the quadrant of $
\tilde{c}_{\tilde{p}}$ it is $ -1 \leq sin\theta_{\tilde{p}} \leq
0$. Replacing this result in (26) and giving, as before, $ \xi_2 -
\xi_1 \neq 0$ a small imaginary part for $ -\frac{\pi}{2} <
\theta_{\tilde{p}} < 0$ yields

\begin{equation}
lim_{|\tilde{p}| \rightarrow \infty}
I_{\tilde{c}_{\tilde{p}}}^{\theta_{\tilde{p}}} = 0~~~~;~~~~ x_2 \neq
x_1
\end{equation}

The integral over $ \tilde{p}$ in (18), now taken along the segment
of $ \tilde{c}_{\tilde{p}}$ which lies on the imaginary axis,
corresponds to $ \theta_{\tilde{p}} = - \frac{\pi}{2}$. Again, in
view of (19), (20) and (21) that integral is

$$
I_{\tilde{c}_{\tilde{p}}}^{-\frac{\pi}{2}} = \int_{\infty}^{\epsilon
\rightarrow
0}d|\tilde{p}|(-i)\frac{(-1)^{-i|\tilde{p}|}}{-isinh{(\pi|\tilde{p}|)}}
e^{i\frac{|\tilde{p}|}{\beta}(\xi_2 -
\xi_1)}e^{i\frac{\pi}{2\beta}|\tilde{p}|(\zeta_2 -
\zeta_1)}e^{i\frac{\pi}{\beta}u_0[|\tilde{p}|](\zeta_2 -
\zeta_1)}\frac{e^{i\frac{3\pi}{4\beta}i(\zeta_2 -
\zeta_1)}}{i\sqrt{\zeta_1\zeta_2}}\times
$$

\begin{equation}
\int_{0}^{\infty}dw\frac{e^{-i\frac{\pi}{\beta}w(\zeta_2 -
\zeta_1)}}{-\pi^2(u_0^2[|\tilde{p}|] + w^2) + 4(l^2 + l + 1) +
2\pi^2u_0[|\tilde{p}|]w}
\end{equation}
where - as is evident in Fig.4 - $ \epsilon << 1$ corresponds to the
radius of the circular quadrant centred at zero.

Replacing

$$
\frac{(-1)^{-i|\tilde{p}|}}{sinh{(\pi|\tilde{p}|)}} =
2\frac{e^{\pi|\tilde{p}|}}{e^{\pi|\tilde{p}|} - e^{-\pi|\tilde{p}|}}
= \frac{2}{1 - e^{-2\pi|\tilde{p}|}}
$$
in (28) yields

$$
I_{\tilde{c}_{\tilde{p}}}^{-\frac{\pi}{2}} =
-2i\int_{\infty}^{\epsilon \rightarrow 0}d|\tilde{p}|\frac{1}{1 -
e^{-2\pi|\tilde{p}|}} e^{i\frac{|\tilde{p}|}{\beta}(\xi_2 -
\xi_1)}e^{i\frac{\pi}{2\beta}|\tilde{p}|(\zeta_2 -
\zeta_1)}e^{i\frac{\pi}{\beta}u_0[|\tilde{p}|](\zeta_2 -
\zeta_1)}\frac{e^{-\frac{3\pi}{4\beta}(\zeta_2 -
\zeta_1)}}{\sqrt{\zeta_1\zeta_2}}\times
$$

\begin{equation}
\int_{0}^{\infty}dw\frac{e^{-i\frac{\pi}{\beta}w(\zeta_2 -
\zeta_1)}}{-\pi^2(u_0^2[|\tilde{p}|] + w^2) + 4(l^2 + l + 1) +
2\pi^2u_0[|\tilde{p}|]w}
\end{equation}

At once, since $ \tilde{c}_{\tilde{p}}$ in Fig.4 does not include
any poles Cauchy's theorem yields

$$
\oint_{\tilde{c}_{\tilde{p}}}d\tilde{p}\frac{(-1)^{\tilde{p}}}{sin{(\pi\tilde{p})}}
e^{i\frac{\tilde{p}}{\beta}(\tau_2 -
\tau_1)}e^{i\frac{\pi}{2\beta}\tilde{p}(\rho_2 -
\rho_1)}e^{i\frac{\pi}{\beta}u_0[\tilde{p}](\rho_2 -
\rho_1)}\frac{e^{i\frac{3\pi}{4\beta}(\rho_2 -
\rho_1)}}{\sqrt{\rho_1\rho_2}}\times
$$

\begin{equation}
\int_{0}^{\infty}dw\frac{e^{-\frac{\pi}{\beta}w(\rho_2 -
\rho_1)}}{\pi^2(u_0^2[\tilde{p}] - w^2) + 4(l^2 + l + 1) +
2i\pi^2u_0[\tilde{p}]w} = 0
\end{equation}
for the stated integral in (18) taken along that infinite contour.
Consequently, in view of (27), (29) and (30) it follows that

$$
I_{c'_{\tilde{p}}}^{+} = -2i\int_{\epsilon \rightarrow
0}^{\infty}d|\tilde{p}|\frac{1}{1- e^{-2\pi|\tilde{p}|}}
e^{i\frac{|\tilde{p}|}{\beta}(\xi_2 -
\xi_1)}e^{i\frac{\pi}{2\beta}|\tilde{p}|(\zeta_2 -
\zeta_1)}e^{i\frac{\pi}{\beta}u_0[|\tilde{p}|](\zeta_2 -
\zeta_1)}\frac{e^{-\frac{3\pi}{4\beta}(\zeta_2 -
\zeta_1)}}{\sqrt{\zeta_1\zeta_2}}\times
$$

\begin{equation}
\int_{0}^{\infty}dw\frac{e^{-i\frac{\pi}{\beta}w(\zeta_2 -
\zeta_1)}}{-\pi^2(u_0^2[|\tilde{p}|] + w^2) + 4(l^2 + l + 1) +
2\pi^2u_0[|\tilde{p}|]w}
\end{equation}

Finally, following always the same procedure in the $
\tilde{p}$-complex plane it is readily seen that at $ |\tilde{p}|
\rightarrow 0$ the contribution stemming from the integral along the
stated small semi-circle in Fig.3 is \footnote{For notational
convenience the index $ \tilde{p}$ in $ \theta_{\tilde{p}}$ will,
henceforth, be dropped.}

$$
I_{c'_{\tilde{p}}}^{sc} =
\frac{i}{\pi}\frac{1}{\sqrt{\zeta_1\zeta_2}}\int_{\frac{\pi}{2}}^{\frac{3\pi}{2}}d\theta
e^{i\theta}e^{i\frac{\pi}{\beta}\big{[}u_{0}[0] +
\frac{3}{4}e^{-i\theta}\big{]}(\zeta_2 - \zeta_1)}\times
$$

\begin{equation}
\int_0^{\infty}dw\frac{e^{-\frac{\pi}{\beta}we^{-i\theta}(\zeta_2 -
\zeta_1)}}{\pi^2(u_0^2[0]e^{2i\theta} - w^2) + 4(l^2 + l + 1) +
2i\pi^2u_0[0]e^{i\theta}w}
\end{equation}

As stated, the integral over $ \tilde{p}$ in (18) taken along the
infinite contour $ c'_{\tilde{p}}$ of Fig.3 is equal to $
I_{c'_{\tilde{p}}}^{+} + I_{c'_{\tilde{p}}}^{\frac{\pi}{2}} +
I_{c'_{\tilde{p}}}^{sc}$. Hence, as a result of (31), (25) and (32)
it is

$$
\frac{1}{2\beta}\oint_{c_{\tilde{p}}}d\tilde{p}\frac{(-1)^{\tilde{p}}}{sin{(\pi\tilde{p})}}
e^{i\frac{\tilde{p}}{\beta}(\tau_2 -
\tau_1)}e^{i\frac{\pi}{2\beta}\tilde{p}(\rho_2 -
\rho_1)}e^{i\frac{\pi}{\beta}u_0[\tilde{p}](\rho_2 -
\rho_1)}\frac{e^{i\frac{3\pi}{4\beta}(\rho_2 -
\rho_1)}}{\sqrt{\rho_1\rho_2}}\times
$$

$$
\int_{0}^{\infty}dw\frac{e^{-\frac{\pi}{\beta}w(\rho_2 -
\rho_1)}}{\pi^2(u_0^2[\tilde{p}] - w^2) + 4(l^2 + l + 1) +
2i\pi^2u_0[\tilde{p}]w} =
$$

$$
-\frac{i}{\beta}\frac{e^{-\frac{3\pi}{4\beta}(\zeta_2 -
\zeta_1)}}{\sqrt{\zeta_1\zeta_2}}\int_{\epsilon \rightarrow
0}^{\infty}d|\tilde{p}|\frac{1}{1 - e^{-2\pi|\tilde{p}|}}
e^{i\frac{|\tilde{p}|}{\beta}(\xi_2 -
\xi_1)}e^{i\frac{\pi}{\beta}\big{[}\frac{|\tilde{p}|}{2} +
u_0[|\tilde{p}|]\big{]}(\zeta_2 - \zeta_1)}\times
$$

$$
\int_{0}^{\infty}dw\frac{e^{-i\frac{\pi}{\beta}w(\zeta_2 -
\zeta_1)}}{-\pi^2(u_0^2[|\tilde{p}|] + w^2) + 4(l^2 + l + 1) +
2\pi^2u_0[|\tilde{p}|]w}
$$

$$
-\frac{i}{\beta}\frac{e^{\frac{3\pi}{4\beta}(\zeta_2 -
\zeta_1)}}{\sqrt{\zeta_1\zeta_2}}\int_{\epsilon \rightarrow
0}^{\infty}d|\tilde{p}|\frac{1}{1 - e^{-2\pi|\tilde{p}|}}
e^{i\frac{|\tilde{p}|}{\beta}(\xi_2 -
\xi_1)}e^{i\frac{\pi}{\beta}\big{[}\frac{|\tilde{p}|}{2} +
u_0[|\tilde{p}|]\big{]}(\zeta_2 - \zeta_1)}\times
$$

$$
\int_{0}^{\infty}dw\frac{e^{i\frac{\pi}{\beta}w(\zeta_2 -
\zeta_1)}}{-\pi^2(u_0^2[|\tilde{p}|] + w^2) + 4(l^2 + l + 1) -
2\pi^2u_0[|\tilde{p}|]w}
$$

$$
+ \frac{i}{\beta}
\frac{1}{2\pi}\frac{e^{i\frac{\pi}{\beta}u_0[0](\zeta_2 -
\zeta_1)}}{\sqrt{\zeta_1\zeta_2}}\int_{\frac{\pi}{2}}^{\frac{3\pi}{2}}d\theta
e^{i\theta}e^{i\frac{3\pi}{4\beta}e^{-i\theta}(\zeta_2 -
\zeta_1)}\times
$$

\begin{equation}
\int_0^{\infty}dw\frac{e^{-\frac{\pi}{\beta}we^{-i\theta}(\zeta_2 -
\zeta_1)}}{\pi^2(u_0^2[0]e^{2i\theta} - w^2) + 4(l^2 + l + 1) +
2i\pi^2u_0[0]e^{i\theta}w}
\end{equation}

This is the analytical extension of (11). It corresponds to
expressing the latter along the imaginary axis of the transfer-space
variables $ u$ and $ p$ respectively. Since $ \frac{\xi_2 -
\xi_1}{\beta}$ has the same angular significance as $ \frac{\tau_2 -
\tau_1}{\beta}$ each of the first two terms on the right side of
(33) is a multiple-valued function of $ \xi_2 - \xi_1$ on account of
the presence of $ |\tilde{p}| \in R$ in the corresponding
exponential. However, as already stated, the sum-total of the three
terms on the right side of (33) - being an integral representation
of (11) - is always a single-valued function of the temporal
coordinate difference $ \xi_2 - \xi_1$.

Through (21) the rotation of $ p$ to purely imaginary values implies
manifestly that $ \rho = \pm i|\rho_{ext}|$ with $ |\rho_{ext}| =
\zeta$ being the magnitude of the radial variable $ \rho_{ext}$ in
the exterior region of the Schwarzschild black-hole geometry - that
is, of the radial variable in (2). This result is inherent in (33).
At once, (2) implies that for $ r \leq 2M$ the radial variable $
\rho_{int}$ in the interior region satisfies $ \rho_{int} = \pm
i|\rho_{int}|$. Consequently, the radial variable $ \rho$ in (33) is
in coincidence with $ \rho_{int}$. This result, namely that

\begin{equation}
\rho_{int} = \pm i|\rho_{ext}| ~~~ ; ~~~ |\rho_{int}| = |\rho_{ext}|
\end{equation}
expresses the fact that the right side of (33) corresponds to an
analytical extension of (11) in the interior region of the
Schwarzschild black-hole geometry. At once, at the coincidence
space-time limit $ x_2 \rightarrow x_1$ each of the first two terms
on the right side of (33) explicitly features the desired
logarithmic divergence while remaining manifestly convergent away
from that limit. Although at $ x_2 \rightarrow x_1$ the expression
in (33) itself is formally indeterminate it is clear that the
independent presence of the stated logarithmic divergence in the
first two terms signifies a logarithmic divergence for the entire
expression in (33). This fact shall be rigorously established in
section V for the entire Green function in four dimensions.
Consequently, the expression on the right side of (33) is, indeed,
the physically valid analytical extension of (11) in the interior
region of the Schwarzschild black hole. The crucial aspect in its
derivation has been the demand that the analytical extension be such
as to preserve the logarithmic divergence of (11) at each stage of
the calculation.

It should also be remarked that, although to each value of $ \rho$
in (2) there corresponds a value of $ |\rho_{int}|$, there is an
infinite range of values of $ |\rho_{int}|$ which do not correspond
to any value of $ \rho$ in the exterior region. This is readily
understood since in the exterior region of the black hole it is $ 0
\leq \rho \leq \beta$ whereas in the interior region it is $ 0 <
|\rho_{int}| < \infty$. Consequently, in terms of the Schwarzschild
radial variable $ r$, the maximum possible range of validity of this
analytical extension of $ D(x_2 - x_1)$ in the interior region of
the black hole is $ (0, M)$. Of course, as will be established in
due course, the actual range of validity in the interior region is
smaller than that since $ D(x_2 - x_1)$ is, in the exterior region,
valid for a subset of values of $ r \in [0, \infty)$.

The analytical extension of (12) will be obtained through the same
procedure so that the analysis leading to the final result will be
concise.

Upon extending $ u \in R$ to $ \tilde{u} \in C$ Cauchy's theorem
implies

\begin{equation}
\oint_{e_{\tilde{u}}}d\tilde{u}\frac{e^{-i\frac{\pi}{\beta}\tilde{u}(\rho_2
- \rho_1)}}{\pi^2\tilde{u}^2 + 4(l^2 + l + 1)} = 0
\end{equation}
with $ e_{\tilde{u}}$ being the contour in Fig.5.

\begin{figure}[h]
\centering\epsfig{figure=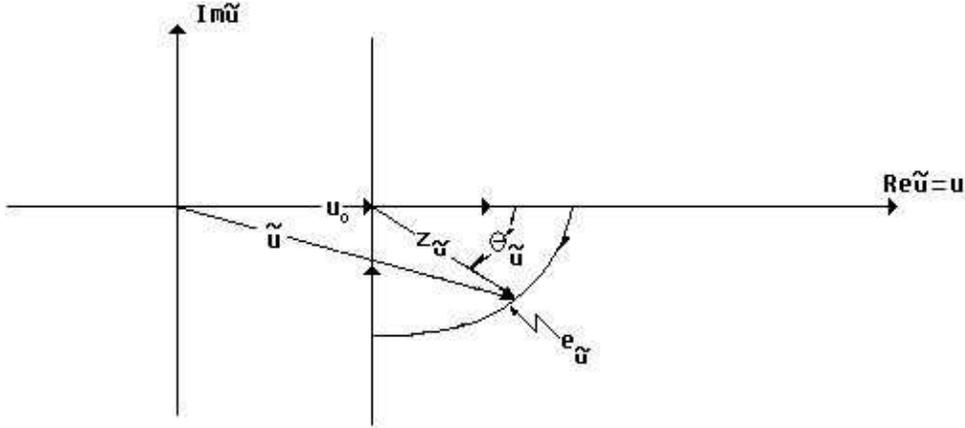, height =85mm,width=141mm}
\caption{Integration over the infinite contour $ e_{\tilde{u}}$ in
the $ \tilde{u}$-complex plane. The choice of $ e_{\tilde{u}}$ is
dictated by the demand for vanishing contribution along the infinite
quadrant.}
\end{figure}

Setting $ \tilde{u} - u_0 = z_{\tilde{u}}$ it is on the large
quadrant $ e_{\tilde{u}}^{\theta'}$ of $ e_{\tilde{u}}$

$$
I_{e_{\tilde{u}}^{\theta'}} = ie^{-i\frac{\pi}{\beta}u_0[p](\rho_2 -
\rho_1)}\int_0^{-\frac{\pi}{2}}d\theta'_{\tilde{u}}|z_{\tilde{u}}|
e^{i\theta_{\tilde{u}}'}\frac{e^{-i\frac{\pi}{\beta}|z_{\tilde{u}}|
cos\theta_{\tilde{u}}'(\rho_2 -
\rho_1)}e^{\frac{\pi}{\beta}|z_{\tilde{u}}|sin\theta_{\tilde{u}}'(\rho_2
- \rho_1)}}{\pi^2(z_{\tilde{u}} + u_0)^2 + 4(l^2 + l + 1)}
$$
so that letting $ |\tilde{u}| \rightarrow \infty$ while keeping $
u_0$ at a fixed value it is

\begin{equation}
lim_{|z_{\tilde{u}}| \rightarrow \infty}I_{e_{\tilde{u}}^{\theta'}}
= 0
\end{equation}

The contribution along the imaginary segment of the infinite contour
$ e_{\tilde{u}}$ is

$$
I_{e_{\tilde{u}}^{-\pi/2}} = \int_{u_0 -
i\infty}^{u_0[p]}d\tilde{u}\frac{e^{-i\frac{\pi}{\beta}\tilde{u}(\rho_2
- \rho_1)}}{\pi^2\tilde{u}^2 + 4(l^2 + l + 1)} =
e^{-i\frac{\pi}{\beta}u_0[p](\rho_2 - \rho_1)}
\int_{-i\infty}^0dz_{\tilde{u}}\frac{e^{-i\frac{\pi}{\beta}z_{\tilde{u}}(\rho_2
- \rho_1)}}{\pi^2(z_{\tilde{u}} + u_0)^2 + 4(l^2 + l + 1)}
$$
with $ |z_{\tilde{u}}| = w$ it is

\begin{equation}
I_{e_{\tilde{u}}^{-\pi/2}} = e^{-i\frac{\pi}{\beta}u_0[p](\rho_2 -
\rho_1)}
\int_{\infty}^0dwe^{-i\frac{\pi}{2}}\frac{e^{-\frac{\pi}{\beta}w(\rho_2
- \rho_1)}}{\pi^2(u_0^2 - w^2) + 4(l^2 + l + 1) - 2i\pi^2u_0w}
\end{equation}

As a consequence of (36) and (37) the statement in (35) implies

\begin{equation}
\int_{u_0[p]}^{\infty}du\frac{e^{-i\frac{\pi}{\beta}u(\rho_2 -
\rho_1)}}{\pi^2u^2 + 4(l^2 + l + 1)} =
-ie^{-i\frac{\pi}{\beta}u_0[p](\rho_2 - \rho_1)}
\int_{0}^{\infty}dw\frac{e^{-\frac{\pi}{\beta}w(\rho_2 -
\rho_1)}}{\pi^2(u_0^2 - w^2) + 4(l^2 + l + 1) - 2i\pi^2u_0w}
\end{equation}

In view of this result (12) becomes

$$
-i\frac{e^{-i\frac{3\pi}{4\beta}(\rho_2 -
\rho_1)}}{\beta\sqrt{\rho_1\rho_2}}\sum_{p =
0}^{\infty}e^{i\frac{p}{\beta}(\tau_2 -
\tau_1)}e^{-i\frac{\pi}{2\beta}p(\rho_2 -
\rho_1)}e^{-i\frac{\pi}{\beta}u_0[p](\rho_2 - \rho_1)}\times
$$

\begin{equation}
\int_{0}^{\infty}dw\frac{e^{-\frac{\pi}{\beta}w(\rho_2 -
\rho_1)}}{\pi^2(u_0^2[p] - w^2) + 4(l^2 + l + 1) - 2i\pi^2u_0[p]w}
\end{equation}

As was the case with (11) it is necessary to, analytically, extend
the transfer-space variable $ p$ and convert the infinite series
over $ p$ in (39) to a contour integral. The residue theorem then
implies that (39) can be expressed as

$$
-\frac{i}{\beta}\frac{1}{2i}\oint_{c'_{\tilde{p}}}d\tilde{p}\frac{(-1)^{\tilde{p}}}{sin{(\pi\tilde{p})}}
e^{i\frac{\tilde{p}}{\beta}(\tau_2 -
\tau_1)}e^{-i\frac{\pi}{2\beta}\tilde{p}(\rho_2 -
\rho_1)}e^{-i\frac{\pi}{\beta}u_0[\tilde{p}](\rho_2 -
\rho_1)}\frac{e^{-i\frac{3\pi}{4\beta}(\rho_2 -
\rho_1)}}{\sqrt{\rho_1\rho_2}}\times
$$

\begin{equation}
\int_{0}^{\infty}dw\frac{e^{-\frac{\pi}{\beta}w(\rho_2 -
\rho_1)}}{\pi^2(u_0^2[\tilde{p}] - w^2) + 4(l^2 + l + 1) -
2i\pi^2u_0[\tilde{p}]w}
\end{equation}
along the infinite contour $ c'_{\tilde{p}}$ of Fig.3. Replacing
(19), (20) and (21) in the integral over $ \tilde{p}$ in (40) taken
along the large quadrant of $ c'_{\tilde{p}}$ and following the
procedure relevant to (22) and (23) causes the stated integral to
vanish at $ |\tilde{p}| \rightarrow \infty$.

The integral over $ \tilde{p}$ in (40), now taken along that segment
of $ c'_{\tilde{p}}$ which lies on the imaginary axis, corresponds
to $ \theta_{\tilde{p}} = \frac{\pi}{2}$. In the context of (19),
(20) and (21) that integral is

$$
\tilde{I}_{c'_{\tilde{p}}}^{\frac{\pi}{2}} =
2i\int_{\infty}^{\epsilon \rightarrow 0}d|\tilde{p}|\frac{1}{1 -
e^{-2\pi|\tilde{p}|}}e^{i\frac{|\tilde{p}|}{\beta}(\xi_2 -
\xi_1)}e^{-i\frac{\pi}{2\beta}|\tilde{p}|(\zeta_2 -
\zeta_1)}e^{-i\frac{\pi}{\beta}u_0[|\tilde{p}|](\zeta_2 -
\zeta_1)}\frac{e^{-\frac{3\pi}{4\beta}(\zeta_2 -
\zeta_1)}}{\sqrt{\zeta_1\zeta_2}}\times
$$

\begin{equation}
\int_{0}^{\infty}dw\frac{e^{i\frac{\pi}{\beta}w(\zeta_2 -
\zeta_1)}}{-\pi^2(u_0^2[|\tilde{p}|] + w^2) + 4(l^2 + l + 1) +
2\pi^2u_0[|\tilde{p}|]w}
\end{equation}

In effect, the integral over $ \tilde{p}$ in (40) is equal to $
\tilde{I}_{c'_{\tilde{p}}}^{+} +
\tilde{I}_{c'_{\tilde{p}}}^{\frac{\pi}{2}} +
\tilde{I}_{c'_{\tilde{p}}}^{sc}$ - with the notation relevant to the
first and the third term having the same significance as $
I_{c'_{\tilde{p}}}^{+}$ and $ I_{c'_{\tilde{p}}}^{sc}$ respectively.

Again, the evaluation of the Cauchy principal value $
\tilde{I}_{c'_{\tilde{p}}}^{+}$ necessitates the corresponding
evaluation of the integral over $ \tilde{p}$ in (40) along the
infinite contour $ \tilde{c}_{\tilde{p}}$ of Fig.4. Replacing (19),
(20) and (21) in the integral over $ \tilde{p}$ in (40) taken along
the large quadrant of $ \tilde{c}_{\tilde{p}}$ and following the
procedure relevant to (26) and (27) causes the stated integral to
vanish at $ |\tilde{p}| \rightarrow \infty$.

The integral over $ \tilde{p}$ in (40), now taken along that segment
of $ \tilde{c}_{\tilde{p}}$ which lies on the imaginary axis,
corresponds to $ \theta_{\tilde{p}} = - \frac{\pi}{2}$. Always in
view of (19), (20) and (21) that integral is

$$
\tilde{I}_{\tilde{c}_{\tilde{p}}}^{- \frac{\pi}{2}} = -
2i\int_{\infty}^{\epsilon \rightarrow 0}d|\tilde{p}|\frac{1}{1 -
e^{-2\pi|\tilde{p}|}} e^{i\frac{|\tilde{p}|}{\beta}(\xi_2 -
\xi_1)}e^{-i\frac{\pi}{2\beta}|\tilde{p}|(\zeta_2 -
\zeta_1)}e^{-i\frac{\pi}{\beta}u_0[|\tilde{p}|](\zeta_2 -
\zeta_1)}\frac{e^{\frac{3\pi}{4\beta}(\zeta_2 -
\zeta_1)}}{\sqrt{\zeta_1\zeta_2}}\times
$$

\begin{equation}
\int_{0}^{\infty}dw\frac{e^{- i\frac{\pi}{\beta}w(\zeta_2 -
\zeta_1)}}{-\pi^2(u_0^2[|\tilde{p}|] + w^2) + 4(l^2 + l + 1) -
2\pi^2u_0[|\tilde{p}|]w}
\end{equation}

At once, Cauchy's theorem yields a vanishing result for the integral
in (40) taken along $ \tilde{c}_{\tilde{p}}$. Consequently, (42)
implies

$$
\tilde{I}_{c'_{\tilde{p}}}^{+} = -2i\int_{\epsilon \rightarrow
0}^{\infty}d|\tilde{p}|\frac{1}{1 - e^{-2\pi|\tilde{p}|}}
e^{i\frac{|\tilde{p}|}{\beta}(\xi_2 -
\xi_1)}e^{-i\frac{\pi}{2\beta}|\tilde{p}|(\zeta_2 -
\zeta_1)}e^{-i\frac{\pi}{\beta}u_0[|\tilde{p}|](\zeta_2 -
\zeta_1)}\frac{e^{\frac{3\pi}{4\beta}(\zeta_2 -
\zeta_1)}}{\sqrt{\zeta_1\zeta_2}}\times
$$

\begin{equation}
\int_{0}^{\infty}dw\frac{e^{- i\frac{\pi}{\beta}w(\zeta_2 -
\zeta_1)}}{-\pi^2(u_0^2[|\tilde{p}|] + w^2) + 4(l^2 + l + 1) -
2\pi^2u_0[|\tilde{p}|]w}
\end{equation}

Finally, at $ |\tilde{p}| \rightarrow 0$ the contribution $
\tilde{I}_{c'_{\tilde{p}}}^{sc}$ stemming from the integral along
the stated small semi-circle in Fig.3 is

$$
\tilde{I}_{c'_{\tilde{p}}}^{sc} =
\frac{i}{\pi}\frac{1}{\sqrt{\zeta_1\zeta_2}}\int_{\frac{\pi}{2}}^{\frac{3\pi}{2}}d\theta
e^{i\theta}e^{-i\frac{\pi}{\beta}\big{[}u_{0}[0] +
\frac{3}{4}e^{-i\theta}\big{]}(\zeta_2 - \zeta_1)}\times
$$

\begin{equation}
\int_0^{\infty}dw\frac{e^{-\frac{\pi}{\beta}we^{-i\theta}(\zeta_2 -
\zeta_1)}}{\pi^2(u_0^2[0]e^{2i\theta} - w^2) + 4(l^2 + l + 1) -
2i\pi^2u_0[0]e^{i\theta}w}
\end{equation}

As stated, the integral over $ \tilde{p}$ in (40) taken along the
infinite contour $ c'_{\tilde{p}}$ of Fig.3 is equal to $
\tilde{I}_{c'_{\tilde{p}}}^{+} +
\tilde{I}_{c'_{\tilde{p}}}^{\frac{\pi}{2}} +
\tilde{I}_{c'_{\tilde{p}}}^{sc}$. Hence, as a result of (43), (41)
and (44) it is

$$
-\frac{1}{2\beta}\oint_{c'_{\tilde{p}}}d\tilde{p}\frac{(-1)^{\tilde{p}}}{sin{(\pi\tilde{p})}}
e^{i\frac{\tilde{p}}{\beta}(\tau_2 -
\tau_1)}e^{-i\frac{\pi}{2\beta}\tilde{p}(\rho_2 -
\rho_1)}e^{-i\frac{\pi}{\beta}u_0[\tilde{p}](\rho_2 -
\rho_1)}\frac{e^{-i\frac{3\pi}{4\beta}(\rho_2 -
\rho_1)}}{\sqrt{\rho_1\rho_2}}\times
$$

$$
\int_{0}^{\infty}dw\frac{e^{-\frac{\pi}{\beta}w(\rho_2 -
\rho_1)}}{\pi^2(u_0^2[\tilde{p}] - w^2) + 4(l^2 + l + 1) -
2i\pi^2u_0[\tilde{p}]w} =
$$

$$
\frac{i}{\beta}\frac{e^{\frac{3\pi}{4\beta}(\zeta_2 -
\zeta_1)}}{\sqrt{\zeta_1\zeta_2}}\int_{\epsilon \rightarrow
0}^{\infty}d|\tilde{p}|\frac{1}{1 - e^{-2\pi|\tilde{p}|}}
e^{i\frac{|\tilde{p}|}{\beta}(\xi_2 -
\xi_1)}e^{-i\frac{\pi}{\beta}\big{[}\frac{|\tilde{p}|}{2} +
u_0[|\tilde{p}|]\big{]}(\zeta_2 - \zeta_1)}\times
$$

$$
\int_{0}^{\infty}dw\frac{e^{- i\frac{\pi}{\beta}w(\zeta_2 -
\zeta_1)}}{-\pi^2(u_0^2[|\tilde{p}|] + w^2) + 4(l^2 + l + 1) -
2\pi^2u_0[|\tilde{p}|]w}
$$

$$
+ \frac{i}{\beta}\frac{e^{-\frac{3\pi}{4\beta}(\zeta_2 -
\zeta_1)}}{\sqrt{\zeta_1\zeta_2}}\int_{\epsilon \rightarrow
0}^{\infty}d|\tilde{p}|\frac{1}{1 - e^{-2\pi|\tilde{p}|}}
e^{i\frac{|\tilde{p}|}{\beta}(\xi_2 -
\xi_1)}e^{-i\frac{\pi}{\beta}\big{[}\frac{|\tilde{p}|}{2} +
u_0[|\tilde{p}|]\big{]}(\zeta_2 - \zeta_1)}\times
$$

$$
\int_{0}^{\infty}dw\frac{e^{i\frac{\pi}{\beta}w(\zeta_2 -
\zeta_1)}}{-\pi^2(u_0^2[|\tilde{p}|] + w^2) + 4(l^2 + l + 1) +
2\pi^2u_0[|\tilde{p}|]w}
$$

$$
- \frac{i}{\beta}
\frac{1}{2\pi}\frac{e^{-i\frac{\pi}{\beta}u_0[0](\zeta_2 -
\zeta_1)}}{\sqrt{\zeta_1\zeta_2}}\int_{\frac{\pi}{2}}^{\frac{3\pi}{2}}d\theta
e^{i\theta}e^{-i\frac{3\pi}{4\beta}e^{-i\theta}(\zeta_2 -
\zeta_1)}\times
$$

\begin{equation}
\int_0^{\infty}dw\frac{e^{-\frac{\pi}{\beta}we^{-i\theta}(\zeta_2 -
\zeta_1)}}{\pi^2(u_0^2[0]e^{2i\theta} - w^2) + 4(l^2 + l + 1) -
2i\pi^2u_0[0]e^{i\theta}w}
\end{equation}

This is the analytical extension of (12). The analysis as to the
physical significance of (33) applies, of course, identically also
to (45). It should be remarked, in addition, that the last term in
(33) and the corresponding last term in (45) appear to be somewhat
peculiar as they do not feature any time dependence. However, the
presence of these two terms in the final result of the analytical
extension should not be surprising. By the analytical procedure in
the $ \tilde{p}$-complex plane it is clear that they correspond to
the $ p = 0$ terms in (11) and (12) respectively. What may come as a
surprise, nevertheless, is that these two ``innocuous-looking" terms
will prove to be the very cause of the radiation of particles by the
Schwarzschild black hole, as will be seen in section VI.

In the interior of the Schwarzschild black hole the additive result
of (33) and (45) is, in $ 1 + 1$ dimensions, the physically valid
analytical extension of the singular part $ D_{as}(x_2 - x_1)$ of
the Green function in (6).

{\bf IV. The Analytical Extension - Boundary Part}

The boundary part $ D_{b}(x_2 - x_1)$ of the Green function in (6) -
that is, the part which remains finite at $ x_2 \rightarrow x_1$ and
enforces the boundary condition \cite{George}

$$
D(x_2 - x_1)_{|\rho_2 = \beta \bigvee \rho_1 = \beta} = 0
$$
is

$$
D_{b}(x_2 - x_1) = -
\frac{2}{\beta^{\frac{3}{2}}}\frac{1}{\sqrt{\rho_1}}\times
$$

\begin{equation}
\sum_{l = 0} ^{\infty}\sum_{m = -l}^{l}\sum_{p =
0}^{\infty}\int_{u_0'[p]}^{\infty}du\frac{cos[\frac{\pi}{4\beta}(4u
+ 2p + 3)(\beta - \rho_1)]}{\pi^2u^2 + 4(l^2 + l +
1)}\frac{J_p(\frac{2i}{\beta}\sqrt{l^2 + l +
1}\rho_2)}{J_p(2i\sqrt{l^2 + l + 1})}e^{i\frac{p}{\beta}(\tau_2 -
\tau_1)}Y_{lm}(\theta_2, \phi_2)Y_{lm}^*(\theta_1, \phi_1) ~~~;~~
\end{equation}
the radial-temporal sector of which is the sum of

\begin{equation}
- \frac{1}{\beta^{\frac{3}{2}}}\frac{e^{i\frac{3\pi}{4\beta}(\beta -
\rho_1)}}{\sqrt{\rho_1}}\sum_{p =
0}^{\infty}e^{i\frac{p}{\beta}(\tau_2 -
\tau_1)}e^{i\frac{\pi}{2\beta}p(\beta -
\rho_1)}\frac{J_p(\frac{2i}{\beta}\sqrt{l^2 + l +
1}\rho_2)}{J_p(2i\sqrt{l^2 + l +
1})}\int_{u_0'[p]}^{\infty}du\frac{e^{i\frac{\pi}{\beta}u(\beta -
\rho_1)}}{\pi^2u^2 + 4(l^2 + l + 1)}
\end{equation}
and

\begin{equation}
- \frac{1}{\beta^{\frac{3}{2}}}\frac{e^{-i\frac{3\pi}{4\beta}(\beta
- \rho_1)}}{\sqrt{\rho_1}}\sum_{p =
0}^{\infty}e^{i\frac{p}{\beta}(\tau_2 -
\tau_1)}e^{-i\frac{\pi}{2\beta}p(\beta -
\rho_1)}\frac{J_p(\frac{2i}{\beta}\sqrt{l^2 + l +
1}\rho_2)}{J_p(2i\sqrt{l^2 + l +
1})}\int_{u_0'[p]}^{\infty}du\frac{e^{-i\frac{\pi}{\beta}u(\beta -
\rho_1)}}{\pi^2u^2 + 4(l^2 + l + 1)}
\end{equation}

The expressions in (47) and (48) are valid in the exterior region of
the Schwarzschild black-hole space-time and, as before, the
replacement of $ r \geq 2M$ by $ 0 \leq r < 2M$ causes either one of
them to diverge sharply for arbitrary space-time separations. It
would appear, for that matter, that the analytical procedure which
established a convergent expression for the singular part of the
propagator in the interior region can also establish a convergent
expression for the propagator's boundary part. In the present case,
however, there is an additional issue which stems from the fact that
for fixed $ z \in C$ the Bessel functions of the first and second
kind, $ J_{\nu}(z)$ and $ I_{\nu}(z)$ respectively, have an
essential singularity at $ |\nu| \rightarrow \infty$ \cite{Olver}.
In effect, in order to implement the stated analytical procedure it
must first be established that, upon replacing (16) with $ \rho_2 =
\beta$ in (47), the meromorphic \footnote{It is meromorphic in the
sense that all branch cuts exist exclusively in the complexified
coordinate domain. There are no cuts in the $ \tilde{p}$-plane. This
point is further elaborated below.} function which constitutes the
integrand in

$$
- \frac{i}{\beta^{\frac{3}{2}}}\frac{1}{2i} \oint
d\tilde{p}\frac{(-1)^{\tilde{p}}}{sin(\pi\tilde{p})}
e^{i\frac{\tilde{p}}{\beta}(\tau_2 -
\tau_1)}e^{i\frac{\pi}{2\beta}\tilde{p}(\beta -
\rho_1)}e^{i\frac{\pi}{\beta}u_0'[\tilde{p}](\beta -
\rho_1)}\frac{e^{i\frac{3\pi}{4\beta}(\beta -
\rho_1)}}{\sqrt{\rho_1}}\frac{J_{\tilde{p}}(\frac{2i}{\beta}\sqrt{l^2
+ l + 1}\rho_2)}{J_{\tilde{p}}(2i\sqrt{l^2 + l + 1})}\times
$$

\begin{equation}
\int_{0}^{\infty}dw\frac{e^{-\frac{\pi}{\beta}w(\beta -
\rho_1)}}{\pi^2(u_0'^2[\tilde{p}] - w^2) + 4(l^2 + l + 1) +
2i\pi^2u_0'[\tilde{p}]w}
\end{equation}
taken along the infinite contours $ c'_{\tilde{p}}$ and $
\tilde{c}_{\tilde{p}}$ in Fig.3 and Fig.4 respectively contains no
essential singularities also at $ |\tilde{p}| \rightarrow \infty$.
In order to explore the behaviour of the integrand at $ |\tilde{p}|
\rightarrow \infty$ use must be made of the representation

\begin{equation}
J_{\nu}(z) = (\frac{z}{2})^{\nu}\sum_{s =
0}^{\infty}\frac{(-1)^s(\frac{z^2}{4})^s}{s!\Gamma(\nu + s + 1)} ~~~
; ~~~ z \in C, ~~~ \nu \in C
\end{equation}
which implies that, for $ |\tilde{p}| >> 1$, the ratio of the two
Bessel functions in (49) behaves as

\begin{equation}
\frac{J_{\tilde{p}}(\frac{2i}{\beta}\sqrt{l^2 + l +
1}\rho_2)}{J_{\tilde{p}}(2i\sqrt{l^2 + l + 1})} \approx
e^{\tilde{p}ln\frac{\rho_2}{\beta}}[1 + O(\frac{1}{\tilde{p}})]
\end{equation}

As was the case with the singular part it can be seen again that
(47) admits several analytical extensions depending on the manner in
which $ (-1)^{\tilde{p}}$ is expressed in (49). The condition which
this time selects a unique, physically valid analytical extension
is, of course, that the boundary part - and consequently the
analytical extension of (47) and (48) independently - be finite for
all values of $ \xi_2 - \xi_1$ and $ \zeta_2 -\zeta_1$. As a
consequence, any extension which does not meet that requirement must
be ruled out. At once, consistency with this requirement imposes the
additional demand that the analytical extension of (47) and,
independently, that of (48) yield a finite result for all values of
$ \xi_{2,1}$ and $ \zeta_{2,1}$ at each stage of the calculation.

As a consequence of this demand it is $ (-1)^{\tilde{p}} =
(e^{i\pi})^{\tilde{p}}$, if the integral in (49) is taken along $
c'_{\tilde{p}}$. In that case the product between (51) and the
asymptotic expression of $
\frac{(-1)^{\tilde{p}}}{sin(\pi\tilde{p})}$ is

\begin{equation}
2e^{|\tilde{p}|\big{(}cos\theta ln\frac{\zeta_2}{\beta} + (\theta -
2\pi)sin\theta\big{)}}e^{i|\tilde{p}|\big{(}sin\theta
ln\frac{\zeta_2}{\beta} - (\theta - 2\pi)cos\theta\big{)}}[1 +
O(\frac{1}{|\tilde{p}|})]
\end{equation}
In view of the fact that $ \zeta < \beta$ and that, in this case, it
is $ 0 \leq \theta \leq \frac{\pi}{2}$ this product causes the
integral in (49) to vanish exponentially as $ |\tilde{p}|
\rightarrow \infty$.

Likewise, if the integrand in (49) is taken along $
\tilde{c}_{\tilde{p}}$ the demand that the analytical extension of
(47) yield a finite result for all values of $ \xi_{2,1}$ and $
\zeta_{2,1}$ at each stage of the calculation implies that $
(-1)^{\tilde{p}} = (e^{-i\pi})^{\tilde{p}}$. In that case the
product between (51) and the asymptotic expression of $
\frac{(-1)^{\tilde{p}}}{sin(\pi\tilde{p})}$ is

\begin{equation}
2e^{|\tilde{p}|\big{(}cos\theta ln\frac{\zeta_2}{\beta} + (\theta +
2\pi)sin\theta\big{)}}e^{i|\tilde{p}|\big{(}sin\theta
ln\frac{\zeta_2}{\beta} - (\theta + 2\pi)cos\theta\big{)}}[1 +
O(\frac{1}{|\tilde{p}|})]
\end{equation}
and causes the integral in (49) to vanish as $ |\tilde{p}|
\rightarrow \infty$ for all values of $ \theta \in (0,
-\frac{\pi}{2}]$.

In effect, the analytical procedure which yielded the extension of $
D_{as}(x_2 - x_1)$ in (33) and (45) can also be applied to (49) to
the same purpose. Before advancing that procedure, however, it is
important to note that - as was also the case in (18) and (40) - for
each value of $ \tilde{p} \in C$ the integrand of the contour
integral in (49) is a multiple-valued function of $ \tau_2 - \tau_1$
although the contour integral itself is a single-valued function of
that temporal difference. Unlike (18) and (40), however, the
integrand featured in (49) is - for each value of $ \tilde{p}$ -
also a multiple-valued function of $ \rho_2$. This is a direct
consequence of the fact that the Bessel functions of the first and
second kind, $ J_{\nu}(z)$ and $ I_{\nu}(z)$ respectively, are
multiple-valued functions in the $ z$-complex plane unless $ \nu \in
Z$ \cite{Olver} \footnote{This situation is reminiscent of that in
\cite{JenCan} and \cite{CanJen} in which the unknown functions of
the radial coordinate have been shown to feature branch cuts in the
corresponding radial variable.}. Despite that fact it follows again
from the character of (49) as an integral representation of (47)
that the contour integral in (49) is, itself, also a single-valued
function of $ \rho_2$. At once, the contour integral in (49) is -
despite appearances - a single-valued function of $ \rho_1$. This
follows again from the implicit dependence of $ \rho_1$ on $
\tilde{p}$ through (21) and the fact that (49) is an integral
representation of (47). Consequently, the physical demand that the
propagator be regular for all values of space-time coordinates in
its domain is satisfied in the interior of the Schwarzschild black
hole.

The ensuing analysis follows the same pattern as that which was
applied to the singular part $ D_{as}(x_2 - x_1)$. In view of the
stated fact that the contour integral in (49) vanishes along the
infinite quadrant of $ c'_{\tilde{p}}$ it follows that - in the
context of (19), (20) and (21) - the contribution which that contour
integral receives along that segment of $ c'_{\tilde{p}}$ which lies
on the imaginary axis is

$$
\textbf{I}_{c'_{\tilde{p}}}^{\frac{\pi}{2}} =
\int_{\infty}^{\epsilon \rightarrow 0}d|\tilde{p}|\frac{2}{e^{2\pi
|\tilde{p}|} - 1}e^{i\frac{|\tilde{p}|}{\beta}(\xi_2 -
\xi_1)}e^{i\frac{\pi}{\beta}\big{[}\frac{|\tilde{p}|}{2} +
u'_0[|\tilde{p}|]\big{]}(\beta -
\zeta_1)}\frac{e^{\frac{3\pi}{4\beta}(\beta -
\zeta_1)}}{\frac{\sqrt{2}}{2}(1 - i)\sqrt{\zeta_1}}\times
$$

\begin{equation}
\frac{J_{i|\tilde{p}|}(-\frac{2}{\beta}\sqrt{l^2 + l +
1}\zeta_2)}{J_{i|\tilde{p}|}(-2i\sqrt{l^2 + l + 1})}
\int_{0}^{\infty}dw\frac{e^{i\frac{\pi}{\beta}w(\beta -
\zeta_1)}}{-\pi^2({u'_0}^2[|\tilde{p}|] + w^2) + 4(l^2 + l + 1) -
2\pi^2u'_0[|\tilde{p}|]w}
\end{equation}
In arriving at (54) the identities

\begin{equation}
J_{\nu}(z) = \frac{z^{\nu}}{(iz)^{\nu}}I_{\nu}(iz)
\end{equation}

\begin{equation}
J_{\nu}(iz) = \frac{(iz)^{\nu}}{z^{\nu}}I_{\nu}(z)
\end{equation}

\begin{equation}
I_{\nu}(z) = \frac{z^{\nu}}{(iz)^{\nu}}J_{\nu}(iz)
\end{equation}

\begin{equation}
I_{\nu}(iz) = \frac{(iz)^{\nu}}{z^{\nu}}J_{\nu}(z)
\end{equation}
have also been used. It should be remarked that

\begin{equation}
\frac{J_{i|\tilde{p}|}(-\frac{2}{\beta}\sqrt{l^2 + l +
1}\zeta_2)}{J_{i|\tilde{p}|}(-2i\sqrt{l^2 + l + 1})} =
\frac{J_{i|\tilde{p}|}(\frac{2}{\beta}\sqrt{l^2 + l +
1}\zeta_2)}{J_{i|\tilde{p}|}(2i\sqrt{l^2 + l + 1})}
\end{equation}
the right side of which identity could have been directly obtained
for $ \theta  = \frac{\pi}{2}$ in (49) - that is, without use of the
identities in (55) - (58). The expression in (54) must be understood
as equivalent to that stemming from the right side of (59) as well
as to all other expressions allowed by the identities in (55) -
(58).

An immediate consequence of the expression in (52) is

\begin{equation}
\frac{J_{i|\tilde{p}|}(-\frac{2}{\beta}\sqrt{l^2 + l +
1}\zeta_2)}{J_{i|\tilde{p}|}(-2i\sqrt{l^2 + l + 1})} \approx
e^{\frac{\pi}{2}|\tilde{p}|}e^{i|\tilde{p}|ln\frac{\zeta_2}{\beta}}[1
+ 0(\frac{1}{|\tilde{p}|})]
\end{equation}
as a result of which the expression in (54) vanishes at $
|\tilde{p}| \rightarrow \infty$.

As was the case with the contour integrals over $ \tilde{p}$ in (18)
and (40) the integral over $ \tilde{p}$ in (49) taken along the
infinite contour $ c'_{\tilde{p}}$ is equal to $
\textbf{I}_{c'_{\tilde{p}}}^{+} + \textbf{I}
_{c'_{\tilde{p}}}^{\frac{\pi}{2}} + \textbf{I}
_{c'_{\tilde{p}}}^{SC}$ with $ \textbf{I}_{c'_{\tilde{p}}}^{+}$
being the contribution which that contour integral receives along $
c_{\tilde{p}}^{+}$ and with $ \textbf{I}_{c'_{\tilde{p}}}^{SC}$
being the contribution along the small semi-circle centred at zero
and extending from $ \theta_{\tilde{p}} = \frac{\pi}{2}$ to $
\theta_{\tilde{p}} = \frac{3\pi}{2}$ in Fig.3. In view of the stated
fact that the contour integral in (49) vanishes along the infinite
quadrant of $ \tilde{c}_{\tilde{p}}$ in Fig.4 the Cauchy principal
value $ \textbf{I}_{c'_{\tilde{p}}}^{+}$ can - as before - be
evaluated through use of Cauchy's theorem applied to the integral in
(49) taken along $ \tilde{c}_{\tilde{p}}$.

In the context of (55) to (58) the contribution which the contour
integral in (49) receives along that segment of $
\tilde{c}_{\tilde{p}}$ which lies on the imaginary axis is

$$
\textbf{I}_{\tilde{c}_{\tilde{p}}}^{-\frac{\pi}{2}} =
\int_{\infty}^{\epsilon \rightarrow
0}d|\tilde{p}|\frac{2}{e^{2\pi|\tilde{p}|} -
1}e^{i\frac{|\tilde{p}|}{\beta}(\xi_2 -
\xi_1)}e^{i\frac{\pi}{2\beta}|\tilde{p}|(\beta -
\zeta_1)}e^{i\frac{\pi}{\beta}u'_0[|\tilde{p}|](\beta -
\zeta_1)}\frac{e^{-\frac{3\pi}{4\beta}(\beta -
\zeta_1)}}{\frac{\sqrt{2}}{2}(1 + i)\sqrt{\zeta_1}}\times
$$

\begin{equation}
\frac{J_{-i|\tilde{p}|}(\frac{2}{\beta}\sqrt{l^2 + l +
1}\zeta_2)}{J_{-i|\tilde{p}|}(-2i\sqrt{l^2 + l + 1})}
\int_{0}^{\infty}dw\frac{e^{-i\frac{\pi}{\beta}w(\beta -
\zeta_1)}}{-\pi^2({u'_0}^2[|\tilde{p}|] + w^2) + 4(l^2 + l + 1) +
2\pi^2u'_0[|\tilde{p}|]w}
\end{equation}

Parenthetically, it shall be remarked again that - pursuant to (59)
- one of the several expressions equivalent to (61) can be obtained
from the identity

\begin{equation}
\frac{J_{-i|\tilde{p}|}(\frac{2}{\beta}\sqrt{l^2 + l +
1}\zeta_2)}{J_{-i|\tilde{p}|}(-2i\sqrt{l^2 + l + 1})} =
\frac{J_{-i|\tilde{p}|}(-\frac{2}{\beta}\sqrt{l^2 + l +
1}\zeta_2)}{J_{-i|\tilde{p}|}(2i\sqrt{l^2 + l + 1})}
\end{equation}

An immediate consequence of the expression in (53) is

\begin{equation}
\frac{J_{-i|\tilde{p}|}(\frac{2}{\beta}\sqrt{l^2 + l +
1}\zeta_2)}{J_{-i|\tilde{p}|}(-2i\sqrt{l^2 + l + 1})} \approx
e^{\frac{\pi}{2}|\tilde{p}|}e^{-i|\tilde{p}|ln\frac{\zeta_2}{\beta}}[1
+ 0(\frac{1}{|\tilde{p}|})]
\end{equation}
as a result of which the expression in (61) vanishes at $
|\tilde{p}| \rightarrow \infty$.

Since the function which in (49) is integrated over $ \tilde{p}$ is
regular throughout the region bounded by the infinite contour $
\tilde{c}_{\tilde{p}}$ it follows that

$$
\textbf{I}_{c'_{\tilde{p}}}^{+} = \int_{\epsilon \rightarrow
0}^{\infty}d|\tilde{p}|\frac{2}{e^{2\pi|\tilde{p}|} - 1}
e^{i\frac{|\tilde{p}|}{\beta}(\xi_2 -
\xi_1)}e^{i\frac{\pi}{\beta}\big{[}\frac{|\tilde{p}|}{2} +
u'_0[|\tilde{p}|]\big{]}(\beta -
\zeta_1)}\frac{e^{-\frac{3\pi}{4\beta}(\beta -
\zeta_1)}}{\frac{\sqrt{2}}{2}(1 + i)\sqrt{\zeta_1}}\times
$$

\begin{equation}
\frac{J_{-i|\tilde{p}|}(\frac{2}{\beta}\sqrt{l^2 + l +
1}\zeta_2)}{J_{-i|\tilde{p}|}(-2i\sqrt{l^2 + l + 1})}
\int_{0}^{\infty}dw\frac{e^{-i\frac{\pi}{\beta}w(\beta -
\zeta_1)}}{-\pi^2({u'_0}^2[|\tilde{p}|] + w^2) + 4(l^2 + l + 1) +
2\pi^2u'_0[|\tilde{p}|]w}
\end{equation}

Finally, it can readily be verified that at $ |\tilde{p}|
\rightarrow 0$ the contribution stemming from the integral along the
stated small semi-circle in Fig.3 is

$$
\textbf{I}_{c'_{\tilde{p}}}^{SC} =
\frac{i}{\pi}\frac{1}{\sqrt{\zeta_1}}\int_{\frac{\pi}{2}}^{\frac{3\pi}{2}}d\theta
e^{i\frac{\theta}{2}}e^{i\frac{\pi}{\beta}\big{[}u'_0[0] +
\frac{3}{4}e^{-i\theta}\big{]}(\beta -
\zeta_1)}\frac{I_{0}(\frac{2}{\beta}\sqrt{l^2 + l +
1}e^{-i\theta}\zeta_2)}{I_{0}(2\sqrt{l^2 + l + 1})}\times
$$

\begin{equation}
\int_0^{\infty}dw\frac{e^{-\frac{\pi}{\beta}we^{-i\theta}(\beta -
\zeta_1)}}{\pi^2({u'_0}^2[0]e^{2i\theta} - w^2) + 4(l^2 + l + 1) +
2i\pi^2u'_0[0]e^{i\theta}w}
\end{equation}

As stated, the contour integral over $ \tilde{p}$ in (49) taken
along the infinite contour $ c'_{\tilde{p}}$ in Fig.3 is equal to $
\textbf{I}_{c'_{\tilde{p}}}^{\frac{\pi}{2}} +
\textbf{I}_{c'_{\tilde{p}}}^{+} + \textbf{I}_{c'_{\tilde{p}}}^{SC}$.
Hence, as a result of (54), (64) and (65) it is

$$
- \frac{1}{2\beta^{\frac{3}{2}}}
\oint_{c'_{\tilde{p}}}d\tilde{p}\frac{(-1)^{\tilde{p}}}{sin(\pi\tilde{p})}
e^{i\frac{\tilde{p}}{\beta}(\tau_2 -
\tau_1)}e^{i\frac{\pi}{2\beta}\tilde{p}(\beta -
\rho_1)}e^{i\frac{\pi}{\beta}u_0'[\tilde{p}](\beta -
\rho_1)}\frac{e^{i\frac{3\pi}{4\beta}(\beta -
\rho_1)}}{\sqrt{\rho_1}}\frac{J_{\tilde{p}}(\frac{2i}{\beta}\sqrt{l^2
+ l + 1}\rho_2)}{J_{\tilde{p}}(2i\sqrt{l^2 + l + 1})}\times
$$

$$
\int_{0}^{\infty}dw\frac{e^{-\frac{\pi}{\beta}w(\beta -
\rho_1)}}{\pi^2(u_0'^2[\tilde{p}] - w^2) + 4(l^2 + l + 1) +
2i\pi^2u_0'[\tilde{p}]w} =
$$

$$
+\frac{\sqrt{2}}{\beta^{\frac{3}{2}}}\frac{e^{\frac{3\pi}{4\beta}(\beta
- \zeta_1)}}{(1 - i)\sqrt{\zeta_1}}\int_{\epsilon \rightarrow
0}^{\infty}d|\tilde{p}|\frac{1}{e^{2\pi|\tilde{p}|} -
1}e^{i\frac{|\tilde{p}|}{\beta}(\xi_2 -
\xi_1)}e^{i\frac{\pi}{\beta}\big{[}\frac{|\tilde{p}|}{2} +
u'_0[|\tilde{p}|]\big{]}(\beta - \zeta_1)}\times
$$

$$
\frac{J_{i|\tilde{p}|}(-\frac{2}{\beta}\sqrt{l^2 + l +
1}\zeta_2)}{J_{i|\tilde{p}|}(-2i\sqrt{l^2 + l + 1})}
\int_{0}^{\infty}dw\frac{e^{i\frac{\pi}{\beta}w(\beta -
\zeta_1)}}{-\pi^2({u'_0}^2[|\tilde{p}|] + w^2) + 4(l^2 + l + 1) -
2\pi^2u'_0[|\tilde{p}|]w}
$$

$$
-\frac{\sqrt{2}}{\beta^{\frac{3}{2}}}\frac{e^{-\frac{3\pi}{4\beta}(\beta
- \zeta_1)}}{(1 + i)\sqrt{\zeta_1}}\int_{\epsilon \rightarrow
0}^{\infty}d|\tilde{p}|\frac{1}{e^{2\pi|\tilde{p}|} -
1}e^{i\frac{|\tilde{p}|}{\beta}(\xi_2 -
\xi_1)}e^{i\frac{\pi}{\beta}\big{[}\frac{|\tilde{p}|}{2} +
u'_0[|\tilde{p}|]\big{]}(\beta - \zeta_1)}\times
$$

$$
\frac{J_{-i|\tilde{p}|}(\frac{2}{\beta}\sqrt{l^2 + l +
1}\zeta_2)}{J_{-i|\tilde{p}|}(-2i\sqrt{l^2 + l + 1})}
\int_{0}^{\infty}dw\frac{e^{-i\frac{\pi}{\beta}w(\beta -
\zeta_1)}}{-\pi^2({u'_0}^2[|\tilde{p}|] + w^2) + 4(l^2 + l + 1) +
2\pi^2u'_0[|\tilde{p}|]w}
$$

$$
-\frac{i}{\beta^{\frac{3}{2}}}
\frac{1}{2\pi}\frac{e^{i\frac{\pi}{\beta}u'_0[0](\beta -
\zeta_1)}}{\sqrt{\zeta_1}}\int_{\frac{\pi}{2}}^{\frac{3\pi}{2}}d\theta
e^{i\frac{\theta}{2}}e^{i\frac{3\pi}{4\beta}e^{-i\theta}(\beta -
\zeta_1)}\frac{I_{0}(\frac{2}{\beta}\sqrt{l^2 + l +
1}e^{-i\theta}\zeta_2)}{I_{0}(2\sqrt{l^2 + l + 1})}\times
$$

\begin{equation}
\int_0^{\infty}dw\frac{e^{-\frac{\pi}{\beta}we^{-i\theta}(\beta -
\zeta_1)}}{\pi^2({u'_0}^2[0]e^{2i\theta} - w^2) + 4(l^2 + l + 1) +
2i\pi^2u'_0[0]e^{i\theta}w}
\end{equation}

This is the analytical extension of (47). It is a consequence of the
character of the integrand in (49) as a meromorphic function in the
$ \tilde{p}$-plane containing no essential singularities at infinity
for $ \theta_{\tilde{p}} \in [-\frac{\pi}{2}, \frac{\pi}{2}]$. As
was the case in (33) and (45) each of the first two terms on the
right side of (66) is a multiple-valued function of $ \xi_2 -
\xi_1$. However, as already stated, the sum-total of the three terms
on the right side of (66) - being an integral representation of (47)
- is always a single-valued function of $ \xi_2 - \xi_1$ and of the
radial coordinates $ \rho_2$ and $ \rho_1$.

The analytical extension of (48) follows the same pattern as that of
(47). Replacing (38) with $ \rho_2 = \beta$ in (48) and converting
the infinite series over $ p$ to a contour integral over the contour
$ c'_{\tilde{p}}$ yields

$$
\frac{1}{2\beta^{\frac{3}{2}}}
\oint_{c'_{\tilde{p}}}d\tilde{p}\frac{(-1)^{\tilde{p}}}{sin(\pi\tilde{p})}
e^{i\frac{\tilde{p}}{\beta}(\tau_2 -
\tau_1)}e^{-i\frac{\pi}{2\beta}\tilde{p}(\beta -
\rho_1)}e^{-i\frac{\pi}{\beta}u_0'[\tilde{p}](\beta -
\rho_1)}\frac{e^{-i\frac{3\pi}{4\beta}(\beta -
\rho_1)}}{\sqrt{\rho_1}}\frac{J_{\tilde{p}}(\frac{2i}{\beta}\sqrt{l^2
+ l + 1}\rho_2)}{J_{\tilde{p}}(2i\sqrt{l^2 + l + 1})}\times
$$

$$
\int_{0}^{\infty}dw\frac{e^{-\frac{\pi}{\beta}w(\beta -
\rho_1)}}{\pi^2(u_0'^2[\tilde{p}] - w^2) + 4(l^2 + l + 1) -
2i\pi^2u_0'[\tilde{p}]w} =
$$

$$
-
\frac{\sqrt{2}}{\beta^{\frac{3}{2}}}\frac{e^{-\frac{3\pi}{4\beta}(\beta
- \zeta_1)}}{(1 - i)\sqrt{\zeta_1}}\int_{\epsilon \rightarrow
0}^{\infty}d|\tilde{p}|\frac{1}{e^{2\pi|\tilde{p}|} - 1}
e^{i\frac{|\tilde{p}|}{\beta}(\xi_2 -
\xi_1)}e^{-i\frac{\pi}{\beta}\big{[}\frac{|\tilde{p}|}{2} +
u'_0[|\tilde{p}|]\big{]}(\beta - \zeta_1)}\times
$$

$$
\frac{J_{i|\tilde{p}|}(-\frac{2}{\beta}\sqrt{l^2 + l +
1}\zeta_2)}{J_{i|\tilde{p}|}(-2i\sqrt{l^2 + l + 1})}
\int_{0}^{\infty}dw\frac{e^{i\frac{\pi}{\beta}w(\beta -
\zeta_1)}}{-\pi^2({u'_0}^2[|\tilde{p}|] + w^2) + 4(l^2 + l + 1) +
2\pi^2u'_0[|\tilde{p}|]w}
$$

$$
+
\frac{\sqrt{2}}{\beta^{\frac{3}{2}}}\frac{e^{\frac{3\pi}{4\beta}(\beta
- \zeta_1)}}{(1 + i)\sqrt{\zeta_1}}\int_{\epsilon \rightarrow
0}^{\infty}d|\tilde{p}|\frac{1}{e^{2\pi|\tilde{p}|} - 1}
e^{i\frac{|\tilde{p}|}{\beta}(\xi_2 -
\xi_1)}e^{-i\frac{\pi}{\beta}\big{[}\frac{|\tilde{p}|}{2} +
u'_0[|\tilde{p}|]\big{]}(\beta - \zeta_1)}\times
$$

$$
\frac{J_{-i|\tilde{p}|}(\frac{2}{\beta}\sqrt{l^2 + l +
1}\zeta_2)}{J_{-i|\tilde{p}|}(-2i\sqrt{l^2 + l + 1})}
\int_{0}^{\infty}dw\frac{e^{-i\frac{\pi}{\beta}w(\beta -
\zeta_1)}}{-\pi^2({u'_0}^2[|\tilde{p}|] + w^2) + 4(l^2 + l + 1) -
2\pi^2u'_0[|\tilde{p}|]w}
$$

$$
+ \frac{i}{\beta^{\frac{3}{2}}}
\frac{1}{2\pi}\frac{e^{-i\frac{\pi}{\beta}u'_0[0](\beta -
\zeta_1)}}{\sqrt{\zeta_1}}\int_{\frac{\pi}{2}}^{\frac{3\pi}{2}}d\theta
e^{i\frac{\theta}{2}}e^{-i\frac{3\pi}{4\beta}e^{-i\theta}(\beta -
\zeta_1)}\frac{I_{0}(\frac{2}{\beta}\sqrt{l^2 + l +
1}e^{-i\theta}\zeta_2)}{I_{0}(2\sqrt{l^2 + l + 1})}\times
$$

\begin{equation}
\int_0^{\infty}dw\frac{e^{-\frac{\pi}{\beta}we^{-i\theta}(\beta -
\zeta_1)}}{\pi^2({u'_0}^2[0]e^{2i\theta} - w^2) + 4(l^2 + l + 1) -
2i\pi^2u'_0[0]e^{i\theta}w}
\end{equation}

The analysis as to the physical significance of the singular part $
D_{as}(x_2 - x_1)$ of the Green function in (6) applies identically
also to (66) and to (67). In addition, (66) and (67) are single
valued and manifestly convergent for any values of $ \tau_i \in [0,
8{\pi}M] ~~~;~~~ i = 1,2$ and for any values of $ \rho_i ~~~;~~~ i =
1,2$ in the range within which the present analytical extension of
(6) is a valid approximation to the Hartle-Hawking Euclidean Green
function in the interior of the Schwarzschild black hole. For that
matter, in $ 1 + 1$ dimensions the additive result of (66) and (67)
is, in the interior of the Schwarzschild black hole, the physically
valid analytical extension of the boundary part $ D_b(x_2 - x_1)$ of
the Green function in (6). The crucial aspect in its derivation has
been the demand that the analytical extension of (47) and (48) be
such as to, independently, preserve - at each stage of the
calculation - the convergent character which these expressions have
for all values of $ \xi_{2,1}$ and $ \zeta_{2,1}$.

{\bf V. Scalar Propagation Inside The Schwarzschild Black Hole}

The results hitherto obtained signify in $ 1 + 1$ dimensions a valid
approximation to the Hartle-Hawking Euclidean Green function for the
massless conformal scalar field in the interior of the Schwarzschild
black hole. The associated range of validity will be established in
what follows. Within that range the approximation to the
Hartle-Hawking Euclidean Green function in the interior of the
Schwarzschild black hole in four dimensions will be obtained by the
appropriate replacement of (6) in the context of the preceding
analytical extensions. Specifically, as a result of (33), (45) and
(9) the contribution which the singular part $ D_{as}(x_2 - x_1)$ of
the Euclidean Green function in (6) has to the corresponding
Euclidean Green function in the interior is

$$
D_{as}^{(int)}(x_2 - x_1) =
\frac{i}{\beta}\frac{e^{-\frac{3\pi}{4\beta}(\zeta_2 -
\zeta_1)}}{\sqrt{\zeta_1\zeta_2}}\int_{\epsilon \rightarrow
0}^{\infty}d|\tilde{p}|\frac{1}{1 - e^{-2\pi|\tilde{p}|}}
e^{i\frac{|\tilde{p}|}{\beta}(\xi_2 -
\xi_1)}e^{i\frac{\pi}{\beta}\big{[}\frac{|\tilde{p}|}{2} +
u_0[|\tilde{p}|]\big{]}(\zeta_2 - \zeta_1)}\times
$$

$$
\sum_{l = 0}^{\infty}\sum_{m = -l}^{l}Y_{lm}(i\tilde{\theta}_2,
\phi_2)Y^*_{lm}(i\tilde{\theta}_1, \phi_1)
\int_{0}^{\infty}dw\frac{e^{-i\frac{\pi}{\beta}w(\zeta_2 -
\zeta_1)}}{\pi^2(u_0[|\tilde{p}|] - w)^2 - 4(l^2 + l + 1)}
$$

$$
+\frac{i}{\beta}\frac{e^{\frac{3\pi}{4\beta}(\zeta_2 -
\zeta_1)}}{\sqrt{\zeta_1\zeta_2}}\int_{\epsilon \rightarrow
0}^{\infty}d|\tilde{p}|\frac{1}{1 - e^{-2\pi|\tilde{p}|}}
e^{i\frac{|\tilde{p}|}{\beta}(\xi_2 -
\xi_1)}e^{i\frac{\pi}{\beta}\big{[}\frac{|\tilde{p}|}{2} +
u_0[|\tilde{p}|]\big{]}(\zeta_2 - \zeta_1)}\times
$$

$$
\sum_{l = 0}^{\infty}\sum_{m = -l}^{l}Y_{lm}(i\tilde{\theta}_2,
\phi_2)Y^*_{lm}(i\tilde{\theta}_1, \phi_1)
\int_{0}^{\infty}dw\frac{e^{i\frac{\pi}{\beta}w(\zeta_2 -
\zeta_1)}}{\pi^2(u_0[|\tilde{p}|] + w)^2 - 4(l^2 + l + 1)}
$$

$$
+ \frac{i}{\beta}
\frac{1}{2\pi}\frac{e^{i\frac{\pi}{\beta}u_0[0](\zeta_2 -
\zeta_1)}}{\sqrt{\zeta_1\zeta_2}}\int_{\frac{\pi}{2}}^{\frac{3\pi}{2}}d\theta
e^{i\theta}e^{i\frac{3\pi}{4\beta}e^{-i\theta}(\zeta_2 -
\zeta_1)}\times
$$

$$
\sum_{l = 0}^{\infty}\sum_{m = -l}^{l}Y_{lm}(i\tilde{\theta}_2,
\phi_2)Y^*_{lm}(i\tilde{\theta}_1, \phi_1)
\int_0^{\infty}dw\frac{e^{-\frac{\pi}{\beta}we^{-i\theta}(\zeta_2 -
\zeta_1)}}{\pi^2(u_0[0]e^{i\theta} + iw)^2 + 4(l^2 + l + 1)}
$$

$$
- \frac{i}{\beta}\frac{e^{\frac{3\pi}{4\beta}(\zeta_2 -
\zeta_1)}}{\sqrt{\zeta_1\zeta_2}}\int_{\epsilon \rightarrow
0}^{\infty}d|\tilde{p}|\frac{1}{1 - e^{-2\pi|\tilde{p}|}}
e^{i\frac{|\tilde{p}|}{\beta}(\xi_2 -
\xi_1)}e^{-i\frac{\pi}{\beta}\big{[}\frac{|\tilde{p}|}{2} +
u_0[|\tilde{p}|]\big{]}(\zeta_2 - \zeta_1)}\times
$$

$$
\sum_{l = 0}^{\infty}\sum_{m = -l}^{l}Y_{lm}(i\tilde{\theta}_2,
\phi_2)Y^*_{lm}(i\tilde{\theta}_1, \phi_1)
\int_{0}^{\infty}dw\frac{e^{-i\frac{\pi}{\beta}w(\zeta_2 -
\zeta_1)}}{\pi^2(u_0[|\tilde{p}|] + w)^2 - 4(l^2 + l + 1)}
$$

$$
- \frac{i}{\beta}\frac{e^{-\frac{3\pi}{4\beta}(\zeta_2 -
\zeta_1)}}{\sqrt{\zeta_1\zeta_2}}\int_{\epsilon \rightarrow
0}^{\infty}d|\tilde{p}|\frac{1}{1 - e^{-2\pi|\tilde{p}|}}
e^{i\frac{|\tilde{p}|}{\beta}(\xi_2 -
\xi_1)}e^{-i\frac{\pi}{\beta}\big{[}\frac{|\tilde{p}|}{2} +
u_0[|\tilde{p}|]\big{]}(\zeta_2 - \zeta_1)}\times
$$

$$
\sum_{l = 0}^{\infty}\sum_{m = -l}^{l}Y_{lm}(i\tilde{\theta}_2,
\phi_2)Y^*_{lm}(i\tilde{\theta}_1, \phi_1)
\int_{0}^{\infty}dw\frac{e^{i\frac{\pi}{\beta}w(\zeta_2 -
\zeta_1)}}{\pi^2(u_0[|\tilde{p}|] - w)^2 - 4(l^2 + l + 1)}
$$

$$
- \frac{i}{\beta}
\frac{1}{2\pi}\frac{e^{-i\frac{\pi}{\beta}u_0[0](\zeta_2 -
\zeta_1)}}{\sqrt{\zeta_1\zeta_2}}\int_{\frac{\pi}{2}}^{\frac{3\pi}{2}}d\theta
e^{i\theta}e^{-i\frac{3\pi}{4\beta}e^{-i\theta}(\zeta_2 -
\zeta_1)}\sum_{l = 0}^{\infty}\sum_{m =
-l}^{l}Y_{lm}(i\tilde{\theta}_2, \phi_2)Y^*_{lm}(i\tilde{\theta}_1,
\phi_1)\times
$$

\begin{equation}
\int_0^{\infty}dw\frac{e^{-\frac{\pi}{\beta}we^{-i\theta}(\zeta_2 -
\zeta_1)}}{\pi^2(u_0[0]e^{i\theta} - iw)^2 + 4(l^2 + l + 1)}
\end{equation}

Inherent in (68) is the expected quadratic divergence at $ x_2
\rightarrow x_1$ which will be analyzed in what follows.

Likewise, as a result of (66), (67) and (9) the contribution which
the boundary part $ D_{b}(x_2 - x_1)$ of the Euclidean Green
function in (6) has, in four dimensions, to the corresponding
Euclidean Green function in the interior of the Schwarzschild black
hole is

$$
D_{b}^{(int)}(x_2 - x_1) =  -
\frac{\sqrt{2}}{\beta^{\frac{3}{2}}}\frac{e^{\frac{3\pi}{4\beta}(\beta
- \zeta_1)}}{(1 - i)\sqrt{\zeta_1}}\int_{\epsilon \rightarrow
0}^{\infty}d|\tilde{p}|\frac{1}{e^{2\pi|\tilde{p}|} -
1}e^{i\frac{|\tilde{p}|}{\beta}(\xi_2 -
\xi_1)}e^{i\frac{\pi}{\beta}\big{[}\frac{|\tilde{p}|}{2} +
u'_0[|\tilde{p}|]\big{]}(\beta - \zeta_1)}\times
$$

$$
\sum_{l = 0}^{\infty}\sum_{m = -l}^{l}Y_{lm}(i\tilde{\theta}_2,
\phi_2)Y^*_{lm}(i\tilde{\theta}_1, \phi_1)
\frac{J_{i|\tilde{p}|}(-\frac{2}{\beta}\sqrt{l^2 + l +
1}\zeta_2)}{J_{i|\tilde{p}|}(-2i\sqrt{l^2 + l + 1})}
\int_{0}^{\infty}dw\frac{e^{i\frac{\pi}{\beta}w(\beta -
\zeta_1)}}{\pi^2(u'_0[|\tilde{p}|] + w)^2 - 4(l^2 + l + 1)}
$$

$$
+
\frac{\sqrt{2}}{\beta^{\frac{3}{2}}}\frac{e^{-\frac{3\pi}{4\beta}(\beta
- \zeta_1)}}{(1 + i)\sqrt{\zeta_1}}\int_{\epsilon \rightarrow
0}^{\infty}d|\tilde{p}|\frac{1}{e^{2\pi|\tilde{p}|} -
1}e^{i\frac{|\tilde{p}|}{\beta}(\xi_2 -
\xi_1)}e^{i\frac{\pi}{\beta}\big{[}\frac{|\tilde{p}|}{2} +
u'_0[|\tilde{p}|]\big{]}(\beta - \zeta_1)}\times
$$

$$
\sum_{l = 0}^{\infty}\sum_{m = -l}^{l}Y_{lm}(i\tilde{\theta}_2,
\phi_2)Y^*_{lm}(i\tilde{\theta}_1, \phi_1)
\frac{J_{-i|\tilde{p}|}(\frac{2}{\beta}\sqrt{l^2 + l +
1}\zeta_2)}{J_{-i|\tilde{p}|}(-2i\sqrt{l^2 + l + 1})}
\int_{0}^{\infty}dw\frac{e^{-i\frac{\pi}{\beta}w(\beta -
\zeta_1)}}{\pi^2(u'_0[|\tilde{p}|] - w)^2 - 4(l^2 + l + 1)}
$$

$$
-\frac{i}{\beta^{\frac{3}{2}}}
\frac{1}{2\pi}\frac{e^{i\frac{\pi}{\beta}u'_0[0](\beta -
\zeta_1)}}{\sqrt{\zeta_1}}\int_{\frac{\pi}{2}}^{\frac{3\pi}{2}}d\theta
e^{i\frac{\theta}{2}}e^{i\frac{3\pi}{4\beta}e^{-i\theta}(\beta -
\zeta_1)}\times
$$

$$
\sum_{l = 0}^{\infty}\sum_{m = -l}^{l}Y_{lm}(i\tilde{\theta}_2,
\phi_2)Y^*_{lm}(i\tilde{\theta}_1, \phi_1)
\frac{I_{0}(\frac{2}{\beta}\sqrt{l^2 + l +
1}e^{-i\theta}\zeta_2)}{I_{0}(2\sqrt{l^2 + l + 1})}
\int_0^{\infty}dw\frac{e^{-\frac{\pi}{\beta}we^{-i\theta}(\beta -
\zeta_1)}}{\pi^2(u'_0[0]e^{i\theta} + iw)^2 + 4(l^2 + l + 1)}
$$

$$ +
\frac{\sqrt{2}}{\beta^{\frac{3}{2}}}\frac{e^{-\frac{3\pi}{4\beta}(\beta
- \zeta_1)}}{(1 - i)\sqrt{\zeta_1}}\int_{\epsilon \rightarrow
0}^{\infty}d|\tilde{p}|\frac{1}{e^{2\pi|\tilde{p}|} -
1}e^{i\frac{|\tilde{p}|}{\beta}(\xi_2 -
\xi_1)}e^{-i\frac{\pi}{\beta}\big{[}\frac{|\tilde{p}|}{2} +
u'_0[|\tilde{p}|]\big{]}(\beta - \zeta_1)}\times
$$

$$
\sum_{l = 0}^{\infty}\sum_{m = -l}^{l}Y_{lm}(i\tilde{\theta}_2,
\phi_2)Y^*_{lm}(i\tilde{\theta}_1, \phi_1)
\frac{J_{i|\tilde{p}|}(-\frac{2}{\beta}\sqrt{l^2 + l +
1}\zeta_2)}{J_{i|\tilde{p}|}(-2i\sqrt{l^2 + l + 1})}
\int_{0}^{\infty}dw\frac{e^{i\frac{\pi}{\beta}w(\beta -
\zeta_1)}}{\pi^2(u'_0[|\tilde{p}|] - w)^2 - 4(l^2 + l + 1)}
$$

$$
-
\frac{\sqrt{2}}{\beta^{\frac{3}{2}}}\frac{e^{\frac{3\pi}{4\beta}(\beta
- \zeta_1)}}{(1 + i)\sqrt{\zeta_1}}\int_{\epsilon \rightarrow
0}^{\infty}d|\tilde{p}|\frac{1}{e^{2\pi|\tilde{p}|} -
1}e^{i\frac{|\tilde{p}|}{\beta}(\xi_2 -
\xi_1)}e^{-i\frac{\pi}{\beta}\big{[}\frac{|\tilde{p}|}{2} +
u'_0[|\tilde{p}|]\big{]}(\beta - \zeta_1)}\times
$$

$$
\sum_{l = 0}^{\infty}\sum_{m = -l}^{l}Y_{lm}(i\tilde{\theta}_2,
\phi_2)Y^*_{lm}(i\tilde{\theta}_1, \phi_1)
\frac{J_{-i|\tilde{p}|}(\frac{2}{\beta}\sqrt{l^2 + l +
1}\zeta_2)}{J_{-i|\tilde{p}|}(-2i\sqrt{l^2 + l + 1})}
\int_{0}^{\infty}dw\frac{e^{-i\frac{\pi}{\beta}w(\beta -
\zeta_1)}}{\pi^2(u'_0[|\tilde{p}|] + w)^2 - 4(l^2 + l + 1)}
$$

$$
+ \frac{i}{\beta^{\frac{3}{2}}}
\frac{1}{2\pi}\frac{e^{-i\frac{\pi}{\beta}u'_0[0](\beta -
\zeta_1)}}{\sqrt{\zeta_1}}\int_{\frac{\pi}{2}}^{\frac{3\pi}{2}}d\theta
e^{i\frac{\theta}{2}}e^{-i\frac{3\pi}{4\beta}e^{-i\theta}(\beta -
\zeta_1)}\times
$$

$$
\sum_{l = 0}^{\infty}\sum_{m = -l}^{l}Y_{lm}(i\tilde{\theta}_2,
\phi_2)Y^*_{lm}(i\tilde{\theta}_1, \phi_1)
\frac{I_{0}(\frac{2}{\beta}\sqrt{l^2 + l +
1}e^{-i\theta}\zeta_2)}{I_{0}(2\sqrt{l^2 + l + 1})}\times
$$

\begin{equation}
\int_0^{\infty}dw\frac{e^{-\frac{\pi}{\beta}we^{-i\theta}(\beta -
\zeta_1)}}{\pi^2(u'_0[0]e^{i\theta} - iw)^2 + 4(l^2 + l + 1)}
\end{equation}

The expressions in (68) and (69) constitute the central result of
this project's entire analysis. Away from the coincidence space-time
limit each of these two expressions is a regular function of the
space-time coordinates. In addition, at $ x_2 \rightarrow x_1$ the
expression in (69) remains regular whereas, as is analyzed below,
the expression in (68) features the expected quadratic divergence.
Consequently, within the corresponding range of validity the
approximation to the Hartle-Hawking Euclidean Green function for the
massless conformal scalar field in the interior of the Schwarzschild
black hole emerges as the additive result of (68) and (69). It
should be obvious that, although invariably finite for all values of
$ \zeta_{1,2} > 0$, the two terms which correspond to $ |\tilde{p}|
\rightarrow 0$ in $ D_{as}^{(int)}(x_2 - x_1)$ are necessary for the
single-valued character of that function. The absorption of these
two terms in the invariably finite $ D_{b}^{(int)}(x_2 - x_1)$ would
result in an inconsistent expression for the asymptotic part of the
Green function. Consequently, always within the associated range of
validity, the contribution $ D_{as}^{(int)}(x_2 - x_1)$ cited in
(68) coincides with the singular part of the Hartle-Hawking
Euclidean Green function for the massless conformal scalar field in
the interior of the Schwarzschild black hole. For that matter, the
contribution $ D_{b}^{(int)}(x_2 - x_1)$ cited in (69) coincides,
itself, with the boundary part of that Green function.

The fact that, at the coincidence space-time limit $ x_2 \rightarrow
x_1$, the dominant contribution to $ D_{as}^{(int)}(x_2 - x_1)$
emerges from the ultra-violet domain $ |\tilde{p}| \rightarrow
\infty$ and $ w \rightarrow \infty$ in transfer space implies at
once that, at $ x_2 \rightarrow x_1$, the dominant contribution to
the integral over $ w$ comes from $ w
>> l$ in each term in the series over $ l$. Consequently,
the completeness relation \cite{CourHilb}

\begin{equation}
\sum_{l = 0}^{\infty}\sum_{m = -l}^{l}Y_{lm}(z_2,
\phi_2)Y^*_{lm}(z_1, \phi_1) = \delta(z_2 - z_1)\delta(\phi_2 -
\phi_1) ~~~;~~~ z_i \in C, i = 1,2
\end{equation}
and the expansion

\begin{equation}
\delta(cosh\theta_2 - cosh\theta_1)\delta(\phi_2 - \phi_1) =
\frac{1}{4\pi}\sum_{k = 0}^{\infty}(2k + 1)P_k(cosh\gamma)
\end{equation}
with

$$
cosh\gamma = cosh\theta_2cosh\theta_1 -
sinh\theta_2sinh\theta_1cos(\phi_2 - \phi_1)
$$
imply by dint of power counting that at $ x_2 \rightarrow x_1$ each
of the first two terms on the right side of the expressions in (33)
and (45) respectively are, independently, quadratically divergent.
This fact, however, does not by itself imply the presence of a
quadratic divergence at $ x_2 \rightarrow x_1$ in (68). In fact,
since (33) and (45) are independently defined for all values of $
\zeta_2$ and $ \zeta_1$ within the associated range of validity it
follows that, as a consequence of the entire analytical procedure,
the asymptotic propagator in (68) is formally indeterminate at $ x_2
\rightarrow x_1$. In order to render concrete meaning to the
expression in (68) at $ x_2 \rightarrow x_1$ use must, instead, be
made of the fact that the additive result of (33) and (45) is an
equivalent representation of the expression in (10) for purely
imaginary values of $ \rho_{1,2}$. The significance of this fact was
analyzed in the context of (33). As a direct consequence, (68)
features, at $ x_2 \rightarrow x_1$, the same divergence structure
as that inherent in (10). In view of \cite{George} it is

\begin{equation}
D_{as}^{(int)}(x_2 \rightarrow x_1) = \frac{1}{2\pi^2\beta^2}\sum_{k
= 0}^{\infty}(2k + 1)\Big{[}\sum_{p = 0}^{p_0
>> 1}\frac{1}{v_p(\zeta)} + \frac{1}{c}\sum_{p =
p_0}^{\infty}\frac{1}{p}\Big{]} + F(\zeta)
\end{equation}
with $ c > 1$ and with $ v_p(\zeta) >1$ and $ F(\zeta)$ being
arbitrary, smooth functions which remain finite throughout their
corresponding domains and such that $ v_p(\zeta) \approx cp$, if $ p
>> 1$. The expression in (72) explicitly manifests the quadratic
divergence expected of the asymptotic propagator in four dimensions
at the coincidence space-time limit in the interior of the black
hole. In conformity with (70) and (71) that quadratic divergence
also characterises, at $ x_2 \rightarrow x_1$, the expressions in
(33) and (45) respectively. In real time the same divergence will,
of course, also occur when two distinct points $ x_2$ and $ x_1$ can
be joined by a null geodesic. The quadratic divergence inherent in
(72) is consistent with theoretical expectations of a scalar
propagator on any manifold.

The analysis hitherto accomplished has yielded, in $ 1+1$ and in
four dimensions respectively, a mathematical expression which is a
valid approximation to the Hartle-Hawking Euclidean Green function
for a massless conformal scalar field within a certain range of
values of the radial coordinate $ r$ in the interior of the
Schwarzschild black hole. In order to assess that range of values
within which the additive result $ D^{(int)}(x_2 - x_1)$ of the
expressions in (68) and (69) is the stated valid approximation it
must be recalled that - as stated in Sec.III - in terms of the
Schwarzschild radial variable $ r$ the maximum possible range of
validity for the analytical extension of $ D(x_2 - x_1)$ in (6) in
the interior of the black hole is $ (0, M)$ since it is only that
range which conforms with (34). However, since the condition in (21)
has been imposed only on those values of $ \rho$ for which (6) is
valid the stated range of validity in the interior is necessarily
restricted by (7) and (8) respectively. In effect, as a result of
(34), the range of validity for the analytical extension
accomplished in the additive result of (68) and (69) is determined
by

\begin{equation}
0 < |\rho_{int}|^2 \leq \frac{\beta^2}{100}
\end{equation}
and accordingly, in terms of $ r$, by

\begin{equation}
1.980M \leq r \leq 2M
\end{equation}
This range of values below the event horizon signifies for the
validity of the approximation to the exact propagator the same order
of magnitude as that expressed in (8) above the event horizon.

As was also the case with the conformal scalar propagator derived in
\cite{George} in the exterior region of the Schwarzschild black-hole
geometry the result

\begin{equation}
D^{(int)}(x_2 - x_1) = D_{as}^{(int)}(x_2 - x_1) + D_b^{(int)}(x_2
-x_1)
\end{equation}
- with $ D_{as}^{(int)}(x_2 - x_1)$ and $ D_b^{(int)}(x_2 - x_1)$
respectively given in (68) and (69) - signifies the only
approximation which specifically and without \emph{a priori}
assumptions expresses the massless conformal scalar propagator  in
the Hartle-Hawking vacuum state in terms of its space-time
dependence in a specific segment of the interior region of the
Schwarzschild black-hole geometry. That segment is specified in (73)
and, equivalently, in (74).

{\bf VI. Scalar Propagator and Black-Hole Radiation}

Having obtained the scalar propagator as an explicitly analytic
expression of space-time in the exterior and interior region of the
Schwarzschild black-hole respectively it is desirable to extract its
inherent information as to the vacuum activity in the Hartle-Hawking
state. At first sight it would appear that the expression in (75) is
far too complicated for any calculation in that direction.
Appearances are - nevertheless - deceptive, as I shall demonstrate
in this section. The first comment which must be made in this
respect is that such a complication was expected from the outset.
The exterior geometry is spherically symmetric and, by dint of
Birkhoff's theorem, also static. The time translational invariance
which the exterior region has with respect to the Schwarzschild time
coordinate $ t$ is expressed by its characteristic global time-like
Killing vector $ \frac{\partial}{\partial t}$. In addition, in the
exterior region the vector $ \frac{\partial}{\partial r}$ is
space-like. In the interior region, however, the Killing vector $
\frac{\partial}{\partial t}$ is space-like whereas the vector $
\frac{\partial}{\partial r}$ is time-like. This situation is an
expression of the fact that in the interior region of the
Schwarzschild black hole the roles of time and space are reversed.
As a consequence, the interior space-time geometry is dynamic and
characterised by the absence of spherical symmetry. The strong
contrast between the substantially more complicated expression in
(75) and that in (6) is a consequence of the stated difference
between the exterior region and the interior region of the
Schwarzschild black hole.

The massless conformal scalar propagator in (6) has already yielded
a prediction regarding the nature of quantum propagation in the
exterior region of the Schwarzschild space-time. In \cite{George} it
was shown that the Green function in (6) vanishes at the limit $
\rho_2 \rightarrow 0$ when $ \rho_1 \neq 0$. As stated in the
relevant analysis the vanishing effect expressed by

\begin{equation}
lim_{\rho_2 \rightarrow 0}D(x_2 - x_1) = 0 ~~~~;~~~~ \rho_1 \neq 0
\end{equation}
is a consequence of the causal structure of the Schwarzschild
black-hole space-time. At the semi-classical limit $ \hbar
\rightarrow 0$ all observers in the static region, regardless of
their state of motion or choice of coordinates, agree that no
particle reaches the hole's event horizon within a finite advance of
their proper time. Equivalently, at the semi-classical limit all
observers in the static region agree that the frequency of a
waveform tends to zero in the vicinity of the hole's event horizon.
As a consequence, away from the semi-classical limit the transition
amplitude for quantum propagation specified by one end-point of
propagation being arbitrarily close to the event horizon is also
expected to vanish for all observers in the static region.

A naive approach to the physical essence of (76) could eventuate in
the conclusion that the latter contradicts the production of
particles by the Schwarzschild black hole. For, if the vacuum is
that of the Hartle-Hawking state, how can an observer who
accelerates to remain stationary at a certain value of the
Schwarzschild radial coordinate $ r$ register any particles
propagating from the vicinity of the event horizon when such a
propagation tends to vanish as a consequence of (76)?

The Penrose diagram in Fig.6 describes the essence of the matter.
The production of particles by the Schwarzschild black hole arises
because the presence of the event horizon renders the Hartle-Hawking
vacuum state unstable. The instability of the Hartle-Hawking vacuum
and the associated radiation produced by the Schwarzschild black
hole can be conceptualised by the production and subsequent
annihilation of a virtual particle-antiparticle pair as an
expression of vacuum fluctuations. The presence of an event horizon
allows for the absence of the subsequent annihilation in the event
that one of the two particles has already crossed that horizon.
Since that particle is, in the interior, necessarily in a quantum
state of negative energy with respect to spatial infinity the other
particle can transit to arbitrarily large radial distances and be
detected as an actual particle in a positive energy state by an
observer situated anywhere in the static region of the Schwarzschild
space-time. I should add that this largely heuristic approach to
Hawking radiation does not define the Hartle-Hawking vacuum state.
It is, instead, a consequence of that state's instability.

The static observer whose world line is represented in Fig.6 is
situated at a radial distance $ r'_1$ from the event horizon which
falls within the range stated in (8). The shaded segment of the
Penrose diagram must be replaced by the space-time geometry in the
interior of the spherical distribution of matter which undergoes
complete gravitational collapse. An equivalent description of the
heuristic approach cited above is represented in the Penrose diagram
of Fig.6. A quantum particle emerges at some point $ B$ of the
future singularity, transits as a positive energy state backward in
time across the future event horizon $ H^+$ where it is scattered by
the background curvature at some point $ C$ in the vicinity of the
latter and continues its transition forward in time always as a
positive energy state before it registers its existence in the
detector of the observer at space-time point $ A \equiv (t'_1, r'_1,
\theta'_1, \phi'_1)$.

\begin{figure}[h]
\centering\epsfig{figure=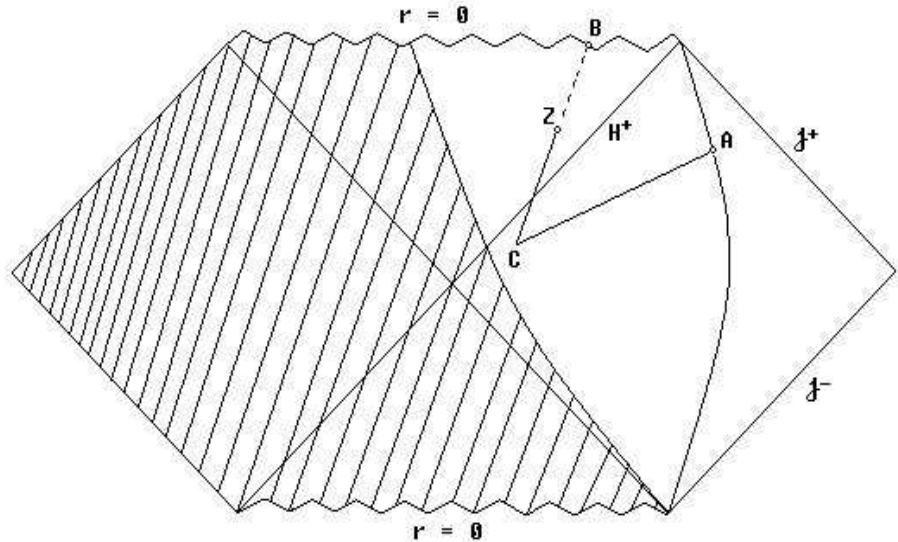, height =105mm,width=141mm}
\caption{Penrose diagram for complete gravitational collapse of a
spherically symmetric distribution of matter. The unshaded segment
represents the geometry outside that distribution. The path $ BCA$
corresponds to particle production by the Schwarzschild black hole
in the Hartle-Hawking vacuum state. The segment $ ZCA$ corresponds
to radial distances from the future event horizon $ H^+$ within
which the propagator in (6) and (75) coincides with the exact
massless scalar propagator. The fixed radial distance of the
observer whose world line is also represented in the diagram lies
within the range of that coincidence.}
\end{figure}

The line $ BCA$ represents the quantum propagation as described
above. The syncopated segment $ BZ$ represents quantum propagation
at radial values below the lower bound $ r_1 = 1.980M$ in (74). The
segment which is depicted by the continuous line from $ Z \equiv
(t_1, r_1, \theta_1, \phi_1)$ to some point on the future event
horizon represents the propagator in the range of validity stated in
(74). Consequently, the segment $ ZCA$ represents the massless
conformal scalar propagator which is described by (75) and (6) in
the interior and the exterior region of the Schwarzschild black hole
respectively.

The probability for the emergence and transition of the quantum
particle from point $ B$ on the future singularity to the
observation point $ A$ is the square of the magnitude of the
transition amplitude from $ B$ to $ A$. That transition amplitude
itself is the functional integral

\begin{equation}
W = \int\mathcal{D}[\Phi] e^{\frac{i}{\hbar}S}
\end{equation}
on condition that the Hartle-Hawking vacuum state of the field $
\Phi(x)$ has been excited only to the stated quantum particle. The
transition amplitude in (77) then is the sum over all paths which
emerge at point $ B$ and eventuate at point $ A$. The action for the
massless conformal scalar field is

\begin{equation}
S = \frac{1}{2}\int d^4x[-g(x)]^{\frac{1}{2}}\big{[}
g^{\mu\nu}(x)\partial_{\mu}\Phi(x)\partial_{\nu}\Phi(x) -
\frac{1}{6} R(x)\Phi^2(x)\big{]}
\end{equation}
with the Schwarzschild metric $g_{\mu\nu}(x)$ expressed in Kruskal
coordinates. In the context of (77) the action functional $ S$
corresponds to the stated set of paths. In the general case which
involves propagation of a field configuration between the future
singularity at $ r = 0$ and the future infinity $ \mathcal{J}^{+}$
the integral which defines the action functional in (78) is taken
over the entire Schwarzschild black-hole space-time (the set of
points which defines the future singularity is not part of
space-time). As the Schwarzschild metric is regular on the event
horizon when expressed in Kruskal coordinates the integral over the
entire space-time is rigorously defined.

The measure $ \mathcal{D}[\Phi]$ in the path integral does not
include paths which emerge from past null infinity $
\mathcal{J^{-}}$ as such paths represent the propagation of incoming
particles from the infinite past. Neither does the measure include
paths which pass through the shaded segment of the Penrose diagram
since, as stated, that region must be replaced by the space-time
geometry in the interior of the spherical distribution of matter
which collapses to a black hole and will not contribute to the
production of particles at late times \cite{HarHawk}.

The path integral which has thus been formally defined can be given
concrete meaning through the analytic continuation of the action to
values which result in integration over an exponentially damped
functional in (77). This can be accomplished by continuing the
Kruskal coordinates to a domain in which the Schwarzschild metric
has signature $ +4$. However, as the objective in this section is to
demonstrate the relation of the propagator in (75) and (6) to
Hawking radiation it is necessary to return to the coordinates in
which the metric receives the expression in (9) and (4) in the
interior and the exterior region respectively, with $ \tau$
identified as an angular variable of period $ 8\pi M$. In the
Euclidean Schwarzschild manifold which emerges as a result of this
arrangement the interior space-time point $ Z$ in Fig.6 corresponds
to the set $ (i\xi_1, i\zeta_1, i\theta_1, \phi_1)$ and the exterior
space-time point $ A$ to $ (\tau'_1, \rho'_1, \theta'_1, \phi'_1)$.
The future event horizon $ (r = 2M)$ in Fig.6 corresponds, on the
Euclidean Schwarzschild manifold, to a two-dimensional sphere of
coordinate values $ (i\zeta = \rho = 0, \theta, \phi)$. The point on
the future event horizon which, in Fig.6, is common to scalar
propagation in the exterior region and in the interior region of the
black hole corresponds to a point which is common to scalar
propagation in (75) and (6) with coordinates $ (i\zeta_2 = \rho_2 =
0, \theta_2, \phi_2)$.

In referring to the Euclidean Schwarzschild manifold it is important
to remember that the latter is the union of the Euclidean
Schwarzschild manifold for $ r \geq 2M$ on which the Green function
in (6) was obtained and of the Euclidean Schwarzschild manifold for
$ r \leq 2M$ on which the Green function in (75) was obtained. As
the Schwarzschild coordinates are singular on the event horizon the
coordinate transformation in (2) implies that - with the angular
dependence on $ \theta$ and $ \phi$ suppressed - the two manifolds
are connected only at the point $ \rho = 0$ - or, equivalently, at $
r = 2M$.

If the vanishing effect expressed in (76) is to be valid in the
event that scalar propagation occurs exclusively in the static
region toward or from the vicinity of the event horizon but not in
the event of scalar propagation which also extends in the interior
region of the Schwarzschild black hole then, necessarily, that
vanishing effect - being a property of scalar propagation
exclusively in the exterior region - must be offset by a property of
scalar propagation which exists exclusively in the interior of the
black hole. In non-rigorous language, we expect the uncertainty
principle - which is heuristically the cause of Hawking radiation in
the Hartle-Hawking vacuum state - to act in such a way as to cancel
the vanishing effect expressed in (76). In what follows I shall show
that this is, indeed, the case.

As, in Fig.6, the Hartle-Hawking vacuum state for the massless
scalar field has been excited to one particle whose position at $
i\xi_1$ and $ \tau'_1$ has been respectively specified to be the
point corresponding to $ Z$ and to $ A$ the amplitude $ <\Phi'_1,
\tau'_1|\Phi_1, i\xi_1>$ for field configuration $ \Phi_1$ at $
i\xi_1$ to propagate, on the Euclidean Schwarzschild manifold, to
field configuration $ \Phi'_1$ at $ \tau'_1$ is identical to the
transition amplitude $ <\rho'_1, \theta'_1, \phi'_1,
\tau'_1|i\zeta_1, i\theta_1, \phi_1, i\xi_1>$. This, in turn,
relates directly to quantum propagation between space-time points $
Z$ and $ A$ on the manifold of Fig.6. On the Euclidean Schwarzschild
manifold for $ r \leq 2M$ the propagator which expresses the
transition amplitude $ <i\zeta_2 = 0, \theta_2, \phi_2,|i\zeta_1,
i\theta_1, \phi_1, i\xi_1>$ is a functional integral over the paths
which connect the point associated with $ Z$ to the common point
which is specified on the two-sphere at $ i\zeta_2 = 0$ $ (r = 2M)$
by $ \theta_2$ and $ \phi_2$. That functional integral is the
Euclidean version of the path integral in (77) with the action in
(78) defined on the Euclidean Schwarzschild manifold for $ r \leq
2M$ and with the Schwarzschild metric $ g_{\mu\nu}(x)$ being
expressed by (9) in the context of (2). It coincides, therefore,
with the propagator in (75) if $ x_1 = (i\zeta_1, i\theta_1, \phi_1,
i\xi_1)$ and $ x_2 = (i\zeta_2 = 0, \theta_2, \phi_2)$. At once, the
propagator which on the Euclidean Schwarzschild manifold for $ r
\geq 2M$ expresses the transition amplitude $ <\rho'_1, \theta'_1,
\phi'_1, \tau'_1|i\zeta_2 = \rho_2 = 0, \theta_2, \phi_2>$ is a
functional integral over the paths which connect the common point on
the two-sphere at $ \rho_2 = 0$ $ (r = 2M)$ to the point associated
with $ A$. That functional integral is, again, the Euclidean version
of the path integral in (77) with the action in (78) defined on the
Euclidean Schwarzschild manifold for $ r \geq 2M$ and with the
Schwarzschild metric being expressed by (4). It coincides,
therefore, with the propagator in (6) if $ x_1 = (\rho'_1,
\theta'_1, \phi'_1, \tau'_1)$ and $ x_2 = (\rho_2 = 0, \theta_2,
\phi_2)$.

This allows for the match of the two transition amplitudes across
the future event horizon $ H^+$. In general, the transition
amplitude $ <\Phi_2, t_2|\Phi_1, t_1>$ is the superposition

\begin{equation}
<\Phi_2, t_2|\Phi_1, t_1> = \int \mathcal{D}\Phi_{t_0}<\Phi_2,
t_2|\Phi_{t_0}, t_0><\Phi_{t_0}, t_0|\Phi_1, t_1>
\end{equation}
with respect to a complete set of modes $ \{|\Phi_{t_0}, t_0>\}$
defined on a space-like hypersurface which corresponds to some $ t_0
\in [t_1, t_2]$ and with $ \mathcal{D}\Phi_{t_0}$ being the
corresponding measure over all field configurations defined on that
hypersurface. A necessary and sufficient condition for (79) is that
the space-time be globally hyperbolic with the space-like
hypersurface being, at once, a Cauchy hypersurface. If the
hypersurface across which the transition amplitudes in (79) are
matched is null, as is the case with the event horizon, then the
situation is substantially more involved. Although, in the case of a
massless field, the past infinity $ \mathcal{J^-}$ is a Cauchy
hypersurface neither is the future infinity $ \mathcal{J^+}$ nor is
the future event horizon $ H^+$ respectively a Cauchy hypersurface
for the collapse space-time in Fig.6. In the case of a massless
field only the union of these two null hypersurfaces is a Cauchy
hypersurface. Consequently, in the case of a massless field, the
modes defined on $ H^+$ do not form a complete set. A complete set
of modes is defined only on the stated union $ H^+ \bigcup
\mathcal{J^+}$ \cite{BirrellDavies}. That precludes the match of the
two transition amplitudes across $ H^+$ in the collapse space-time.
However, as the production of particles by the Schwarzschild black
hole is a consequence of the causal and topological structure of
space-time - as opposed to, a consequence of the space-time geometry
- the transition amplitude between space-time points $ Z$ and $ A$
in the collapse space-time of Fig.6 coincides with the transition
amplitude between the same two points in the maximally extended
Kruskal space-time. Mathematically, this may not be directly obvious
since the measure $ \mathcal{D}\Phi$ in the associated path integral
now includes paths which did not exist in the corresponding measure
of (77). This change is, nevertheless, offset by the presence of
additional field modes in the future event horizon $ H^+$ of the
extended Kruskal manifold. Consequently, quantum propagation can be
examined on the maximally extended Kruskal manifold altogether. The
advantage of such an approach is that the future event horizon $
H^+$ of the extended Kruskal manifold is a Cauchy hypersurface and
admits, for that matter, a complete set of modes.

As a result, the integration which corresponds to that in (79) will
now occur over a complete set of field modes defined on the future
event horizon $ H^+$ of the extended Kruskal manifold. If $ \Phi(x)$
is a massless conformal scalar field then in the Euclidean sector of
the Schwarzschild metric the transition amplitude in (79) becomes

\begin{equation}
<\Phi'_1, \tau'_1|\Phi_1, i\xi_1> = \int \mathcal{D}\Phi_{\tau_2
\rightarrow \infty}\int\mathcal{D}[\Phi]
e^{-\frac{1}{\hbar}I[\tau'_1, \tau_2 \rightarrow
\infty]}\int\mathcal{D}[\Phi] e^{-\frac{1}{\hbar}I[i\xi_2
\rightarrow i\infty, i\xi_1]}
\end{equation}
In (80) $ I[i\xi_2 \rightarrow i\infty, i\xi_1]$ is the action
functional which emerges from the analytical extension of the action
functional in (78) to the Euclidean Schwarzschild manifold for $ r
\leq 2M$ on condition that integration over $ i\xi$ is defined in $
[i\xi_1, i\infty)$. The Euclidean action functional $ I[\tau'_1,
\tau_2 \rightarrow \infty]$ has the corresponding meaning. The
temporal coordinate $ \tau_2 \rightarrow \infty$ labels the event
horizon at $ \rho = 0$ which, in turn, corresponds to $ H^+$ in the
extended Kruskal manifold . For that matter, it also labels all
field configurations on $ H^+$. The notation $ \Phi_{\xi_2
\rightarrow \infty}$ is equivalent.

In the specific situation which is represented in the collapse
Schwarzschild space-time in Fig.6 the scalar field $ \Phi(x)$ has
been excited to one particle and the objective is the match of the
corresponding two scalar propagators in the interior and in the
exterior region across $ H^+$. In view of the preceding analysis,
the physical significance of (80) reduces, in such a physical
context, to

$$
<\rho'_1, \theta'_1, \phi'_1, \tau'_1|i\zeta_1, i\theta_1, \phi_1,
i\xi_1> =
$$

\begin{equation}
\int_{-\pi}^{\pi}d\theta_2\int_{0}^{2\pi}d\phi_2 <\rho'_1,
\theta'_1, \phi'_1, \tau'_1|\rho_2 = 0, \theta_2, \phi_2><i\zeta_2 =
0, \theta_2, \phi_2|i\zeta_1, i\theta_1, \phi_1, i\xi_1>
\end{equation}
with the complete set of modes defined on the future event horizon $
H^+$ of the extended Kruskal manifold.

As stated above the explicit expressions of the two transition
amplitudes featured in the integrand in (81) are accordingly given
by (75) and (6). Since (81) explicitly involves scalar propagation
with one end-point specified arbitrarily close to the event horizon
for both (75) and (6) it is necessary that the asymptotic behaviour
of (75) be also examined before the latter is replaced in (81). The
incipient point for such an examination is the condition $ \frac{\pi
u_{0}[p]}{\beta}\rho_2 >> p$ stated in (6) for scalar propagation in
the exterior region of the Schwarzschild black hole. The latter
implies, through (19) and (21), that in the interior region it is

\begin{equation}
\frac{\pi u_{0}[|\tilde{p}|]}{\beta}\zeta_2 >> |\tilde{p}|
\end{equation}

Clearly, the information which will determine the asymptotic
behaviour of the propagator in (75) is inherent in the expression

\begin{equation}
I[u_0] =
\frac{1}{\sqrt{\zeta_1\zeta_2}}\int_0^{\infty}dw\frac{e^{\pm
i\frac{\pi}{\beta}w(\zeta_2 - \zeta_1)}}{\pi^2(u_0[|\tilde{p}|] \pm
w)^2 - 4(l^2 + l + 1)}
\end{equation}
which is featured in four of the six terms in (68). Since (68) is
defined for all values of $ \zeta_2$ and $ \zeta_1$ within the range
of validity stated in (73) the limit $ \zeta_2 \rightarrow 0$ can be
taken in (68) with $ \zeta_1 \neq 0$. In effect, (82) implies

\begin{equation}
lim_{\zeta_2 \rightarrow
0}\frac{1}{\sqrt{\zeta_1\zeta_2}}\int_0^{\infty}dw\frac{1}{\pi^2(w
\pm u_0)^2 - 4(l^2 + l + 1)} <<
\frac{1}{\beta^2}\frac{1}{|\tilde{p}|^2}\lim_{\Lambda \rightarrow
\infty}\lim_{\zeta_2 \rightarrow
0}\frac{\zeta_2^{\frac{3}{2}}}{\sqrt{\zeta_1}}\int_0^{\Lambda}dw = 0
\end{equation}

The result in (84) implies directly that the real and the imaginary
part of $ I[u_0]$ respectively vanish at $ \zeta_2 \rightarrow 0$ on
condition that $ \zeta_1 \neq 0$. Consequently, at the same limit
each of the stated four terms in (68) vanishes identically. This,
shifts the issue of the scalar propagator's behaviour at $ \zeta_2
\rightarrow 0$ on the remaining two terms in (68) - namely, on the
third and the sixth term.

The third and the sixth term in (68) correspond to $ |\tilde{p}|
\rightarrow 0$. As stated, they are invariably finite for $
\zeta_{1,2} > 0$ and necessary for the single-valued character of $
D_{as}^{(int)}(x_2 - x_1)$. Inspection of the integral over $
\omega$ in these two terms reveals that its mathematical structure
is substantially different from that of the corresponding integral
over $ \omega$ in the previous four terms. The reason is that at $
|\tilde{p}| \rightarrow 0$ (82) allows for two possibilities
regarding $ u_0[0]$. These are respectively, $ u_0[0]
>> \frac{c}{\zeta_2}$ with $ c > 0 ~~~; ~~~ \zeta_2 \neq 0$ and a $
\zeta$-independent $ u_0[0] > 0$. This situation should be
contrasted against the corresponding situation in the exterior
region where the transfer space variable $ p$ receives values from
the discrete set $ 0, 1, 2, ...$. In that case, if the positive $
u_0[0]$ is a $ \rho_2$-independent constant then, at $ \rho_2
\rightarrow 0$ the left side of the condition $ \frac{\pi
u_0[0]}{\beta}\rho_2 >> 0$ lies invariably in a neighbourhood of $ p
= 0$ as is, for example, the subset $ \{0, 1\}$. In addition, this
is always the case however high the value of the constant $ u_0[0]$
may be. Consequently, if $ u_0[0]$ does not depend on $ \rho_2$, the
right side $ \frac{\pi u_0[0]}{\beta}\rho_2$ is not much bigger than
zero and the condition $ \frac{\pi u_0[p]}{\beta}\rho_2 >> p$ is
violated for $ p = 0$, even though the inequality $ \frac{\pi
u_0[0]}{\beta}\rho_2 > 0$ is valid. In effect, in the exterior
region the condition $ \frac{\pi u_0[p]}{\beta}\rho_2 >> p$ implies
exclusively that $ u_0[0] >> \frac{c}{\rho_2}$ with $ c > 1 ~~~;
~~~\rho_2 \neq 0$. Put simply, in the exterior region $ u_0[0]$ can
not - in $ \frac{\pi u_0[0]}{\beta}\rho_2 >> 0$ - be a $
\rho_2$-independent constant because the value $ p = 0$ can always
be replaced by $ p = 0 + 1$ \cite{George}.

Not so in the interior region where the transfer-space variable $
|\tilde{p}|$ receives values in $ [0, +\infty)$. In that case a
positive $ \zeta_2$-independent $ u_0[0]$ is admissible since at $
\zeta_2 \rightarrow 0$ there is at least one neighbourhood of zero
which does not contain $ \frac{\pi u_0[0]}{\beta}\zeta_2$. Both
possibilities concerning $ u_0[0]$ in the interior must be examined.

If $ u_0[0] >> \frac{c}{\zeta_2}$ then at $ \zeta_2 \rightarrow 0$
the third and the sixth term in (68) also vanish identically. If, on
the contrary, $ u_0[0]$ is a positive constant then at $ \zeta_2
\rightarrow 0$ with $ \zeta_1 \neq 0$ the singular part $
D_{as}^{(int)}(x_2 - x_1)$ of the scalar propagator diverges as $
\zeta_2^{-\frac{1}{2}}$. Clearly, an appropriate boundary condition
imposed by the physical context within which this situation occurs
will determine which of these two eventualities is realised. In
order to determine that boundary condition use must be made of the
fact that - as stated in the beginning of this section - in the
interior region of the Schwarzschild black hole the roles of space
and time are reversed. This reversal can be directly elicited by the
change which occurs in the signature of the Schwarzschild metric in
(1) when passing from the region $ r > 2M$ to the region $ r < 2M$.
In the interior region the Schwarzschild radial variable $ r$
becomes time-like. As a result, the decrease of $ r$ in the interior
region labels the advance of time which is, for that matter,
inexorably directed toward the time coordinate $ r = 0$.

As $ r \rightarrow 2M^{-}$ the square of proper time's increment $
ds^2$ in (1) tends to negative infinity. The imaginary character of
the in-falling observer's proper time merely expresses the fact that
his world line is unobservable to all observers in the region $ r >
2M$. At once, the divergence in proper time admits the
interpretation that, expressed in terms of the Schwarzschild
time-like coordinate $ r$, the time registered on the clock of an
observer who has just crossed the event horizon advances at infinite
rate. Consequently, in terms of the Schwarzschild time-like
coordinate $ r$, quantum propagation between two points in the
interior region occurs instantaneously if one of the two points has
been specified arbitrarily close to the event horizon ($ r
\rightarrow 2M^{-}$). As a consequence, described in terms of $ r$
the corresponding transition amplitude must diverge.

The stated divergence in the transition amplitude reflects the
corresponding divergence in the proper time of the in-falling
observer who has just crossed the event horizon. Just like the
latter so the divergence in the transition amplitude is unobservable
in all circumstances. The unobservable divergence in the proper time
itself is not merely a consequence of $ lim_{r \rightarrow 2M^{-}}(1
- \frac{2M}{r}) = 0$ but also of the fact that in the interior
region the situation $ dr = 0$ is impossible. Consequently, although
that divergence is specific to Schwarzschild coordinates it is, at
once, a consequence of the coordinate-independent reversal which
occurs in the interior region to the effect that what was space
effectively becomes time and vice versa. As a result, the
unobservable divergence which occurs in the transition amplitude
away from the coincidence space-time limit is a consequence of the
space-time geometry's causal structure on condition that quantum
propagation is described in terms of Schwarzschild coordinates. Of
course, from the perspective of the in-falling observer himself
nothing unusual happens. Both, the lapse of proper time which his
clock registers and the transition probability for quantum
propagation which he (in principle) observes just after he has
crossed the event horizon remain finite and well defined.

It can readily be seen, for that matter, that the boundary condition
which the physical context imposes on (68) is that the only two
terms which do not vanish at $ \zeta_2 \rightarrow 0$, namely the
third and the sixth term, must diverge. Consequently, in these two
terms $ u_0[0]$ is a positive constant so that these two terms do,
indeed, diverge as $ \zeta_2^{-\frac{1}{2}}$. As a consequence of
this as well as of (84) it follows that, at $ \zeta_2 \rightarrow
0$, the singular part $ D_{as}^{(int)}(x_2 - x_1)$ diverges in its
entirety as $ \zeta_2^{-\frac{1}{2}}$ on condition that $ \zeta_1
\neq 0$. This, in turn, implies that

\begin{equation}
lim_{\zeta_2 \rightarrow 0}D^{(int)}(x_2 - x_1) = \infty ~~~~~ ;
~~~~~ \zeta_1 \neq 0
\end{equation}

In passing, it is worth noting that the condition $ u_0[0] >>
\frac{c}{\rho_2}$ which is valid in the exterior region can also be
established by invoking boundary conditions derived from the
physical context. The boundary condition which, in this case, would
preclude a $ \rho_2$-independent $ u_0[0]$ is that - as stated in
(76) - from the perspective of the observer in the static region the
scalar propagator must vanish if one end-point of propagation is
specified arbitrarily close to the event horizon. However, since the
mathematical context itself already precludes a $
\rho_2$-independent $ u_0[0]$ the mathematical resolution which I
have presented is a superior approach in view of the fact that it
eventuates in the physical interpretation of vanishing propagation
in (76).

The asymptotic behaviour of the propagator expressed, in the
interior and in the exterior region respectively, in (85) and (76)
allows for the evaluation of the transition amplitude in (81). The
latter is

$$
<\rho'_1, \theta'_1, \phi'_1, \tau'_1|i\zeta_1, i\theta_1, \phi_1,
i\xi_1> =
$$

\begin{equation}
\int_{-\pi}^{\pi}d\theta_2\int_{0}^{2\pi}d\phi_2 \big{[}D_{as}(x'_1
- x_2) + D_b(x'_1 - x_2)\big{]}\big{[}D_{as}^{(int)}(x_2 - x_1) +
D_{b}^{(int)}(x_2 - x_1)\big{]}
\end{equation}
with $ x_1 = (i\xi_1, i\zeta_1, i\theta_1, \phi_1)$, $ x_2 = (\rho_2
\rightarrow 0, \theta_2, \phi_2)$ and $ x'_1 = (\tau'_1, \rho'_1,
\theta'_1, \phi'_1)$.

Terms which involve products between a singular part on either side
of the event horizon and a boundary part on the other side do not
contribute to the Green function on the left side of (86). The most
expedient way to establish this in the present context in which
integration occurs over the physically irrelevant angular sector is
to note that whereas $ D_{as}^{(int)}(x_2 - x_1)$ contains a
singularity if the limit $ x_2 \rightarrow x_1$ is taken anywhere
within the range stated in (74) the boundary part $ D_b(x'_1 - x_2)$
remains finite if $ x_2 \rightarrow x'_1$ is taken anywhere within
the range stated in (8). Consequently, the term in (86) which
involves the product $ D_{as}^{(int)}(x_2 - x_1)D_b(x'_1 - x_2)$
contributes neither to the singular part of $ <\rho'_1, \theta'_1,
\phi'_1, \tau'_1|i\zeta_1, i\theta_1, \phi_1, i\xi_1>$ nor to the
boundary part of that Green function. For the same reason the term
in (86) which involves the product $ D_{as}(x'_1 -
x_2)D_b^{(int)}(x_2 - x_1)$ has no contribution to $ <\rho'_1,
\theta'_1, \phi'_1, \tau'_1|i\zeta_1, i\theta_1, \phi_1, i\xi_1>$.
Formally, these two ``cross terms" in (86) cancel against the
corresponding terms which are inherent in $ D_{as}(x'_1 -
x_2)D_{as}^{(int)}(x_2 - x_1)$. Specifically, $ D_{as}(x'_1 - x_2)$
and $ D_{as}^{(int)}(x_2 - x_1)$ can always be expressed as $
\tilde{D}_{as}(x'_1 - x_2) + \tilde{D}_{fin}(x'_1 - x_2)$ and $
\tilde{D}_{as}^{(int)}(x_2 - x_1) + \tilde{D}_{fin}^{(int)}(x_2 -
x_1)$ respectively in such a manner as to enforce $
|\tilde{D}_{as}(x'_1 - x_2)\tilde{D}_{fin}^{(int)}(x_2 - x_1)| =
|D_{as}(x'_1 - x_2)D_b^{(int)}(x_2 - x_1)|$ and $
|\tilde{D}_{as}^{(int)}(x_2 - x_1)\tilde{D}_{fin}(x'_1 - x_2)| =
|D_{as}^{(int)}(x_2 - x_1)D_b(x'_1 - x_2)|$ \cite{George}. The
boundary term $ D_b(x'_1 - x_2)$ has the opposite sign to that of
the finite term $ \tilde{D}_{fin}(x'_1 - x_2)$ in view of the fact
that the former has been designed to enforce the boundary condition
on the Green function in the exterior region. At once, the
analytical extension of that Green function into the interior region
implies that the boundary term $ D_b^{(int)}(x_2 - x_1)$ has the
opposite sign to that of the finite term $
\tilde{D}_{fin}^{(int)}(x_2 - x_1)$.

Consequently, (86) reduces to

$$
<\rho'_1, \theta'_1, \phi'_1, \tau'_1|i\zeta_1, i\theta_1, \phi_1,
i\xi_1> =
$$

\begin{equation}
\int_{-\pi}^{\pi}d\theta_2\int_{0}^{2\pi}d\phi_2 \big{[}D_{as}(x'_1
- x_2)D_{as}^{(int)}(x_2 - x_1)\big{]} +
\int_{-\pi}^{\pi}d\theta_2\int_{0}^{2\pi}d\phi_2 \big{[}D_b(x'_1 -
x_2)D_b^{(int)}(x_2 - x_1)\big{]}
\end{equation}
A crucial aspect of this derivation is that the contribution of $
D_{as}^{(int)}(x_2 - x_1)$ to the first term on the right side of
(87) enters through the two non-singular terms which correspond to $
|\tilde{p}| \rightarrow 0$ - that is, through the third and the
sixth term in (68) - since it is only these two terms which do not
vanish at $ \zeta_2 \rightarrow 0$. Of course, at the coincidence
space-time limit $ x_2 \rightarrow x_1$ in the interior and on the
event horizon it is the remaining four terms in (68) which are
responsible for the quadratic divergence of $ <\rho'_1, \theta'_1,
\phi'_1, \tau'_1|i\zeta_1, i\theta_1, \phi_1, i\xi_1>$.

In the interior vicinity of the event horizon ($ \zeta_2 << 1$) the
cause of the divergence at $ \zeta_2 \rightarrow 0$ in (85) is, as
stated, the factor $ \zeta_2^{-\frac{1}{2}}$ in the third and the
sixth term of (68). In the exterior vicinity of the event horizon ($
\rho_2 << 1$) the singular part $ D_{as}(x'_1 - x_2)$ in (6) behaves
as \cite{George}

$$
lim_{\rho_2 \rightarrow 0}D_{as}(x'_1 - x_2) =
C\frac{1}{2\pi^2}\frac{1}{\beta^2}\times
$$

\begin{equation}
lim_{\rho_2 \rightarrow 0}\frac{\sqrt{\rho_2}}{\sqrt{\rho_1}}\sum_{k
= 0}^{\infty}\sum_{p \neq 0}^{\infty}e^{i\frac{p}{\beta}(\tau_2 -
\tau_1)}\frac{2k + 1}{p}P_k(cos\gamma) ~~~; ~~~ \rho'_1 \neq 0 ~~~;
~~~ C << 1 ~~~~; ~~~~ \tau_2 \rightarrow \infty
\end{equation}
Consequently, in the integrand $ D_{as}(x'_1 -
x_2)D_{as}^{(int)}(x_2 - x_1)$ of the first term in (87), the factor
$ \zeta_2^{-\frac{1}{2}}$ cancels against the factor $
\rho_2^{\frac{1}{2}}$ prior to $ |\rho_2| \rightarrow 0$. As a
result, the transition amplitude $ <\rho'_1, \theta'_1, \phi'_1,
\tau'_1|i\zeta_1, i\theta_1, \phi_1, i\xi_1>$ is a regular and
non-vanishing function of space-time. The ``uncertainty principle"
has canceled the vanishing effect expressed in (76). Within a finite
advance of his proper time $ t$ the observer situated at point $ A$
in Fig.6 registers radiation emitted by the Schwarzschild black hole
- as promised!

{\bf VII. Conclusions and Discussion}

Through an appropriate analytical extension performed on the
massless conformal scalar propagator obtained in \cite{George} the
entire analysis herein eventuated in an analytic expression for that
propagator in the interior region of the Schwarzschild black-hole
geometry. For a specific range of values of the Schwarzschild radial
coordinate that expression is a valid approximation to the exact
massless conformal scalar propagator associated with the
Hartle-Hawking vacuum state in the interior region. Being an
analytic expression the stated propagator is explicit in its
dependence on space-time in the interior region of the Schwarzschild
black hole. This approximation to the exact propagator is,
therefore, unique.

As is also the case with the results in \cite{George} an essential
advantage of the propagator which has been obtained herein is that
its short-distance behaviour and singularity structure are manifest.
Despite its complicated structure this propagator is, for that
matter, especially suited for an analytic evaluation of $
<T_{\mu\nu}>$ and $ <\phi^2(x)>$ as well as for perturbative
calculations in the interior of the black hole. In \cite{George} the
renormalised value of $ <\phi^2(x)>$ was obtained in the
Hartle-Hawking vacuum state in the exterior region of the
Schwarzschild black hole. It would be desirable to demonstrate the
capacity of the propagator obtained herein for actual calculations
in perturbative renormalisation by extending that evaluation to the
evaluation of the renormalised value of $ <\phi^2(x)>$ in the
interior region of the black hole and compare it against the
corresponding results in \cite{CanJen}. Such a task is contingent
upon the consistent formulation of a computer program which will
produce the relevant graphs.

In \cite{George} the range of values stated in (8) was shown to
signify a range of validity for the approximation which corresponds
to several orders of magnitude above the range within which
particles are spontaneously created and back-reaction effects are
pronounced. The relation of the range expressed in (74) to the
physical content in the interior of the black hole as well as the
effect which the causal structure of the interior space-time
geometry has to scalar propagation have been explored in section VI.
The range of validity expressed in (74) is of the same order of
magnitude as that expressed in (8). Just as, for that matter, the
latter signifies an ample range of values of $ r > 2M$ within which
the approximation to the exact propagator is valid \cite{George} it
is reasonable to state that (74) also signifies an ample range of
validity in the interior.

An issue of central importance is the match of the propagator
obtained herein in the interior region with that obtained in
\cite{George} in the exterior region, across the event horizon. That
task was accomplished in section VI. It was shown that, as a result
of that match, the massless scalar propagator which corresponds to
the Hartle-Hawking vacuum state results in particle production by
the Schwarzschild black hole which is registered by static observers
situated within the range of the propagator's validity and - by
extension - by any static observer at any distance from the event
horizon. In order to show that the particle production which this
propagator describes coincides with Hawking radiation it is also
necessary to show that that particle production has a thermal
spectrum which corresponds to the relevant surface-gravity related
temperature. That, in turn, may be established by showing that the
probability for the emission of a massless scalar particle of a
certain energy and the absorption of a massless scalar particle of
the same energy relate to each other in the manner established in
\cite{HarHawk}. Past that point the task will be the exploration of
the relation which higher-order radiative effects in the presence of
self-interactions of the $ \phi^4(x)$-type have to the thermal
character of the black-hole radiation in the Hartle-Hawking vacuum
state. Being especially suited for perturbative calculations the
massless scalar propagator obtained in this project has the
potential to advance such an exploration although, at once, in view
of the highly complicated expression which the massless conformal
scalar propagator receives in the interior region of the
Schwarzschild black hole this may impress as a formidable task.
Nevertheless, the derivation of particle production in section VI is
indicative of the fact that a careful approach to the causal
structure of the Schwarzschild black-hole space-time can
substantially simplify the calculations.

It is my intension to address these issues.

\vspace{0.3in}

{\bf Acknowledgements}

I wish to thank my wife Minnie for her unsolicited patience.

\end{document}